\documentclass[reprint,amsmath,amssymb,aip,eqsecnum,jcp]{revtex4-1}

\usepackage[dvips]{graphicx}
\usepackage{color}
\usepackage{amsmath}
\usepackage{amssymb}
\usepackage{pstricks,pst-grad}
\usepackage{amsbsy}
\usepackage{epic}
\usepackage[normalem]{ulem}

\usepackage[linktoc=all,unicode=true,pdfusetitle,  bookmarks=true,bookmarksnumbered=false,bookmarksopen=false, breaklinks=true,pdfborder={0 0 0},backref=false,colorlinks=true]{hyperref}
\hypersetup{linkcolor=blue,citecolor=red}

\usepackage[hyphenbreaks]{breakurl}

\usepackage[normalem]{ulem}

\usepackage{dcolumn}
\usepackage{bm}

\newcommand{\bq}{\begin{eqnarray}}
\newcommand{\eq}{\end{eqnarray}}
\newcommand{\bqn}{\begin{eqnarray*}}
\newcommand{\eqn}{\end{eqnarray*}}
\newcommand{\rr}{\mathbf{r}}
\newcommand\eff{\text{eff}}
\newcommand\HS{\text{HS}}

\newcommand\spi{\text{sp}}

\newcommand\beq{\begin{equation}}
\newcommand\eeq{\end{equation}}
\newcommand\beqa{\begin{eqnarray}}
\newcommand\eeqa{\end{eqnarray}}

\newcommand{\nn}{\nonumber\\}

\begin{document}

\title{Bridging and depletion mechanisms in colloid-colloid effective interactions: A reentrant phase diagram}
\author{Riccardo Fantoni}
\email{rfantoni@ts.infn.it}
\affiliation{Dipartimento di Scienze Molecolari e Nanosistemi, Universit\`a Ca' Foscari Venezia,
Calle Larga S. Marta DD2137, I-30123 Venezia, Italy}
\author{Achille Giacometti}
\email{achille.giacometti@unive.it}
\author{Andr\'es Santos}
\email{andres@unex.es}
\homepage{http://www.unex.es/eweb/fisteor/andres}
\affiliation{Departamento de F\'{\i}sica and Instituto de Computaci\'on Cient\'ifica Avanzada (ICCAEx), Universidad de Extremadura,
E-06071 Badajoz, Spain}

\date{\today}

\begin{abstract}
A general class of nonadditive sticky-hard-sphere binary mixtures, where small and large spheres represent the solvent and the solute, respectively, is introduced. The solute-solute and solvent-solvent interactions are of hard-sphere type, while the solute-solvent interactions are of sticky-hard-sphere type with tunable degrees of size nonadditivity and stickiness. Two particular and complementary limits are studied  using analytical and semi-analytical tools. The first case is characterized by zero  nonadditivity, lending itself to a Percus--Yevick  approximate solution from which the impact of stickiness on the spinodal curves and on the effective solute-solute potential is analyzed. In the opposite nonadditive case, the solvent-solvent diameter is zero  and the model can then be reckoned as an extension of the well-known Asakura--Oosawa model with additional sticky solute-solvent interaction. This latter model has the property that its exact effective one-component problem involves only solute-solute pair potentials for size ratios such that a solvent particle fits inside the interstitial region of three touching solutes. In particular, we explicitly identify the three competing physical mechanisms (depletion, pulling, and bridging) giving rise to the effective interaction. Some remarks on the phase diagram of these two complementary models are also addressed through the use of the Noro--Frenkel criterion
and a first-order perturbation analysis. Our findings suggest reentrance of the fluid-fluid instability as  solvent density (in the first model) or  adhesion (in the second model) is varied.
Some perspectives in terms of the interpretation of recent experimental studies of microgels adsorbed onto large polystyrene particles are discussed.

\end{abstract}

\maketitle
\section{Introduction}
\label{sec1}
\begin{figure}
\includegraphics[width=8cm]{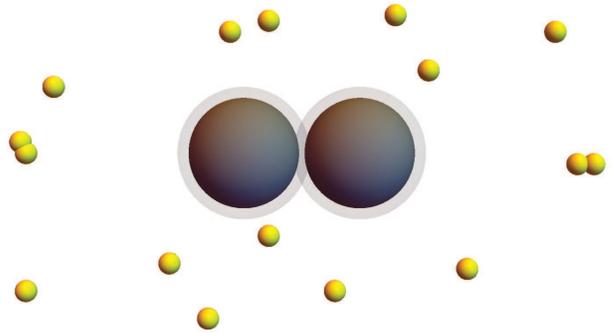}
\caption{{Cartoon of the} {AO} depletion interaction. {The shaded region around each solute represents the volume excluded to the centers of the solvent particles.\protect\cite{note_1}}
\label{fig:fig1}}
\end{figure}

Many years ago, Asakura and Oosawa\cite{Asakura54} {(AO)} provided an explanation of the clustering and gelation phenomenon occurring when
small nonadsorbing polymers, such as polystyrene (PS), were added to
a solution of large spherical colloids, say polymethylmethacrylate (PMMA).
The basic idea is illustrated in Fig.\ \ref{fig:fig1} considering two PMMA colloids, modeled as big spheres, immersed in a fluid formed by a uniform background
(that we will neglect henceforth) as well as by
PS particles, assumed to be small {\emph{noninteracting spheres}} that, however, experience a {hard-sphere (HS)} interaction with the larger ones.
Under these conditions, when the separation between the two large spheres is less than the diameter of the small spheres ({see} Fig.\ \ref{fig:fig1}),  there
is an unbalanced pressure of the ``sea'' of small spheres, providing an entropic gain compared to the case when the separation is large, that can be reckoned as an
{\emph{effective}} attractive interaction driving the clustering of large colloidal spheres.

\begin{figure}
\includegraphics[width=8cm]{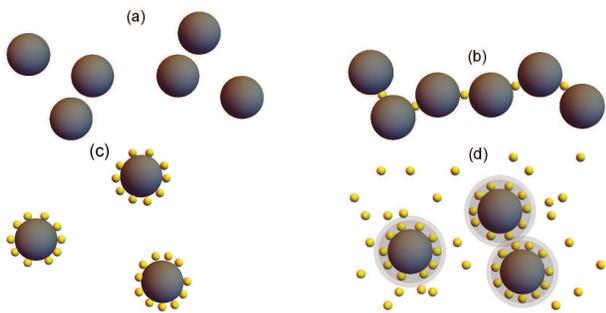}
\caption{Different mechanisms occurring in the presence of a {short-range} attraction between solvent and solute, {as the solvent concentration increases}:
(a) $\eta_s=0$, {HS} behavior; (b) $0<\eta_s< \eta_s^{*}$, the small fraction of solvent particles act as {bridges}
connecting the solute into a cluster; (c) $\eta_s \approx \eta_s^{*}$, {most of the} solute colloids are covered and again behave as {HSs} with an effective diameter $\sigma_s+\sigma_l$; (d) $\eta_s > \eta_s^{*}$, the ``dressed'' solutes feel an effective depletion attraction. {The dark and light shaded regions around the solute particles in panel (d) represent the effective solute size and the  effective volume excluded to the centers of the solvent particles, respectively.}
\label{fig:fig2}}
\end{figure}

In real systems, however, the solvent particles do not always behave as {an} ideal gas {or} interact only sterically. Typically, they experience an additional {short-range}
attraction {(or repulsion)} with {the} solute, usually due to dispersion forces.\cite{Sapir14,Sapir15,BRLRD99,KM13} {The simplest way of accounting for a  short-range solute-solvent attraction is by means of Baxter's sticky-hard-sphere (SHS) model\cite{Baxter68} characterized by a stickiness parameter $\tau_{sl}$}. Both issues {(solvent-solvent repulsion and solute-solvent short-range attraction)} were recently addressed by
two experimental studies\cite{Zhao12a,Zhao12b}
on adsorbing microgels (MG) to large PS latex suspension. In this case,  the expected mechanism will be clearly different, as illustrated
by Fig.\ \ref{fig:fig2}, inspired by a similar figure of Ref.\ \onlinecite{Zhao12a}.

Let $\sigma_l$ and $\sigma_s$ be the diameters of the large and small spheres, respectively, and {suppose} {we} fix the volume fraction {$\eta_l$} of the
large colloidal spheres and gradually increase the volume fraction $\eta_s$ of the small solvent spheres.
In the absence of solvent particles, the solute particles will behave essentially as {HSs}, as depicted
in Fig.\ \ref{fig:fig2}(a). Now imagine {we} gradually add the small solvent particles. Because of the solute-solvent attraction, they will tend to
get {adsorbed on} the surface of the larger particles and mediate an effective attraction between them. This \textit{bridging} mechanism {destabilizes} the
solution as the large colloidal spheres tend to form aggregates, as schematically illustrated in Fig.\ \ref{fig:fig2}(b). The global effect is the
formation of a \textit{gel} phase {caused} by a {free-energy} driven phase separation of the large and small spheres.
As $\eta_s$ increases, solvent particles tend to progressively cover the solute surface, as depicted in Fig.\ \ref{fig:fig2}(c).
We can easily estimate\cite{Fantoni12} the critical value $\eta_s^{*}$ at which all large spheres will be completely covered to be $\eta_s^{*} \approx \eta_l (2\pi/\sqrt{3}) {\sigma_s/\sigma_l}$, as discussed in Appendix \ref{app:simple}.
At this point, all the {solute} colloids {can be ``fully covered''} by {solvent}
particles and they will behave essentially again as {HSs} with {an} effective diameter $\sigma_l+\sigma_s$, with a few additional free solvent particles. This situation is pictured in Fig.\ \ref{fig:fig2}(c). Upon adding further
solvent particles, however,  depletion forces between the small and the covered colloids set in [Fig.\ \ref{fig:fig2}(d)] and phase separation occurs again,
this time entropically rather than {free-energetically}, as in {the case of Fig.\ \ref{fig:fig2}}(b).
{A useful way to represent the phase diagram of such a binary mixture is
through an $(\eta_l, \eta_s)$ diagram at fixed values of size ratio $q=\sigma_s/\sigma_l$ and stickiness
$\tau_{sl}^{-1}$. In this  diagram, there will be geometrically inaccessible
regions, for example for $\eta_s$ or $\eta_l$ larger than $\pi/3\sqrt{2}$, and lines
separating the various phase coexistence regions. The topology of the phase diagram would be controlled by $q$, while  $\eta_s$ would play the role of an
inverse temperature.}

Motivated by these new experimental perspectives, recently
Chen et al.\cite{Chen15} considered a HS-SHS binary mixture
where one can tune the attraction {parameter $\tau_{sl}$} between the unlike spheres, with like spheres only interacting
via {HS} interactions. Note that this is the same model already studied by Fantoni et al.,\cite{Fantoni05}
as well as by other groups.\cite{Barboy79} The  study of Ref.\ \onlinecite{Chen15} provided a well defined framework to rationalize the
experimental results obtained in Refs.\ \onlinecite{Zhao12a} and \onlinecite{Zhao12b}.

In the present work, we will build upon this idea and go further to introduce also an additional ---and, to the best of our knowledge, new---
model that has the interesting feature of including the {standard} AO model\cite{Asakura54,Chen15} as a particular case.
In both cases, we will illustrate how an effective one-component solute-solute {interaction} potential can be obtained and
the merits and {drawbacks} of this procedure.

{Both models can be seen as extreme realizations of a general class of nonadditive sticky-hard-sphere (NASHS) binary mixtures  where the small-small (or solvent-solvent) and large-large (or solute-solute) interactions are of HS type with diameters $\sigma_{ss}$ and $\sigma_{ll}=\sigma_l$, respectively, while the small-large (or solvent-solute) interactions are of SHS type characterized by a cross diameter $\sigma_{sl}=(\sigma_s+\sigma_l)/2=\sigma_l(1+q)/2$ and an inverse stickiness $\tau_{sl}$. Note that here we denote by $\sigma_s{=q\sigma_l}$ the diameter of the small spheres \emph{as seen} by the large ones, while {$\sigma_{ss}$} is the diameter of the small spheres as seen by themselves. Thus,  the nonadditivity of the unlike interactions is monitored by the ratio $\sigma_{ss}/\sigma_s\leq 1$ {(where we have restricted ourselves to zero or positive nonadditivity)}.}
{The NASHS class reduces to the nonadditive hard-sphere (NAHS) class if the solute-solvent stickiness is switched off.}

In the first model {that we will study} {one has} {$\sigma_{ss}/\sigma_s=1$, so that the HS interactions are additive. This model, denoted henceforth as  the additive sticky-hard-sphere (ASHS) model, is the one depicted in Fig.\ \ref{fig:fig2} and considered in Refs.\ \onlinecite{Chen15} and \onlinecite{Fantoni05}. Interestingly, the ASHS model can be solved exactly within the Percus--Yevick (PY) approximation\cite{Fantoni05,PS75,B75,Santos98}} and the instability region in the $(\eta_s,\eta_l)$ plane enclosed by the spinodal line  can be computed.
This will be found to form a closed region, in agreement with previous results.\cite{Chen15}

{The second model represents an extreme case of {positive} nonadditivity, namely $\sigma_{ss}/\sigma_s=0$, i.e., the solvent spheres behave among themselves as an ideal gas. This particular case of the general class of NASHS models reduces to the conventional AO model if the stickiness {is} switched off (i.e., $\tau_{sl}\to\infty$). Because of that, we will term this model as the sticky Asakura--Oosawa (SAO) model. {The need of supplementing the AO model with a short-range solute-solvent attraction has been recognized, for instance, in Ref.\ \onlinecite{BRLRD99}}. While, in contrast to the ASHS model, the SAO model does not allow for an analytical solution in the PY approximation, its associated \emph{effective} solute-solute pair potential can be exactly derived in the semi-grand-canonical ensemble, analogously to the case of the pure AO model.\cite{Dijkstra99,Dijkstra99b,Binder14}. Moreover, and also in analogy with the AO model,\cite{Dijkstra99b,AW14a,Ashton14,SHFS15}  such  a pair potential turns out to be the only one contributing to the exact effective interaction among the solutes if the size ratio $q=\sigma_s/\sigma_l$ is smaller than the threshold value $q_0=2/\sqrt{3}-1\approx 0.1547$.}
A careful comparison between the results of the two models {(ASHS and SAO)}
allows us to pave the way for an improved theoretical understanding of the above experiments.

\begin{figure}
\includegraphics[width=8cm]{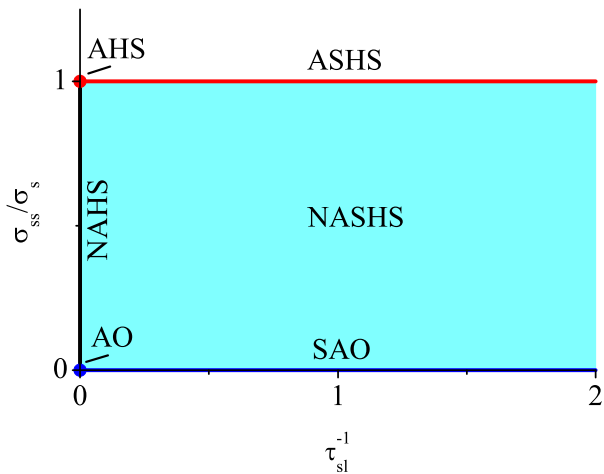}
\caption{{Plane $\sigma_{ss}/\sigma_s$ vs $\tau_{sl}^{-1}$ sketching different models mentioned in the text. The general class of NASHS models include, as limiting cases, ASHS  ($\sigma_{ss}/\sigma_s=1$),  SAO  ($\sigma_{ss}/\sigma_s=0$), and  NAHS ($\tau_{sl}^{-1}=0$). The intersection of the NAHS line with the ASHS and SAO lines define the AHS and AO models, respectively. In this paper, we will be concerned with the ASHS and SAO models.}
\label{fig:fig3}}
\end{figure}

It is interesting to observe that, when the solute-solvent adhesion is set to zero, the model ASHS  reduces to a size-asymmetric additive HS (AHS)
binary mixture, while the  SAO model becomes the original AO model, these two mixtures having quite different critical behaviors upon varying $q$.\cite{Dijkstra99,Dijkstra99b,Zykova10} The
metastable fluid-fluid demixing coexistence, responsible for the
broadening at $\eta_s>0$ of the stable fluid-solid coexistence ($0.492\leq \eta_l\leq 0.543$) for pure
HSs   ($\eta_s\to 0$),\cite{Noya08,Fernandez12} remains always metastable and exists at small enough
$q$ in the AHS case, whereas it becomes stable at large $q$ in the
AO case, where a triple point appears.
{Figure \ref{fig:fig3} sketches  (in the plane $\sigma_{ss}/\sigma_s$ vs $\tau_{sl}^{-1}$) the different models referred to above.}

{The organization of this paper is as follows. Section \ref{sec2} presents the problem of the effective solute interaction mediated by the solvent particles within a general framework. This is followed by Sec.\ \ref{sec:ASHS}, where the PY solution for the ASHS model is exploited to find the spinodal curves of the original mixture and the effective solute-solute pair potential. The exact derivation of the effective potential in the SAO model with a size ratio $q<q_0$ is addressed in Sec.\ \ref{sec:AAO}, its three contributions being clearly identified. Next, the different scenarios for criticality in the ASHS and SAO effective  systems are analyzed via the second virial coefficient and the Noro--Frenkel criterion\cite{Noro00} in Sec.\ \ref{sec:NF}. A more detailed analysis for the SAO model is performed via a first-order perturbation theory in Sec.\ \ref{sec:PTII}. Finally, our findings are discussed and put in perspective in Sec.\ \ref{sec:conclusions}. The most technical details are relegated to four appendixes.}

\section{General framework\label{sec2}}
Consider a colloidal binary mixture of {$N_s$} small (solvent) and {$N_l$} large  (solute)
particles, identified by the coordinates $\{\rr_1^{(s)},\rr_2^{(s)},\ldots,\rr_{N_s}^{(s)}\}$
and  $\{\rr_1^{(l)},\rr_2^{(l)},\ldots,\rr_{N_l}^{(l)}\}$, respectively, {in a volume $V$}.

Assuming pair interactions, {i.e., assuming the particles are nondeformable, nonpolarizable, \ldots,} (see Ref.\ \onlinecite{Menichetti15} for a recent discussion on the reliability of this assumption),
the total potential $U$ can be written as $U=U_{ss}+U_{ll}+U_{sl}$, where
\begin{eqnarray}
U_{ss}&=&{\sum_{i=1}^{N_s-1}\sum_{j=i+1}^{N_s}}\varphi_{ss}(|\rr_i^{(s)}-\rr_j^{(s)}|),\\
U_{ll}&=&{\sum_{i=1}^{N_l-1}\sum_{j=i+1}^{N_l}}\varphi_{ll}(|\rr_i^{(l)}-\rr_j^{(l)}|),\\
U_{sl}&=&{\sum_{i=1}^{N_s}\sum_{j=1}^{N_l}}\varphi_{sl}(|\rr_i^{(s)}-\rr_j^{(l)}|).
\label{general:eq1}
\end{eqnarray}
The canonical free energy $F(N_s,N_l,V,T)$ is then given by
\beq
e^{-\beta F}= \frac{{\Lambda_s^{-3N_s} \Lambda_l^{-3N_l}}}{N_s!N_l! } \int d\mathbf{r}^{N_{s}} \int d\mathbf{r}^{N_{l}}
e^{-\beta\left(U_{ss}+U_{ll}+U_{sl}\right)},
\label{general:eq2}
\eeq
where $\beta=1/k_B T$ ($k_B$ being the Boltzmann constant), $\Lambda_s$ and $\Lambda_l$ are the de Broglie thermal {wavelengths}
associated with the small and large particles, respectively, {and we have used the short-hand notation $d\rr^{N_\alpha}=d\rr_1^{(\alpha)}\cdots d\rr_{N_\alpha}^{(\alpha)}$ with $\alpha=s,l$}.

Following standard prescriptions,\cite{Dijkstra99a,Dijkstra99,Dijkstra99b}
one can  {in principle} trace out all the microscopic degrees of freedom associated
with the solvent particles and recast Eq.\ (\ref{general:eq2}) in
a form of an \textit{effective} one-component system for only the solute particles
with a potential energy $U_{ll}^\eff(\rr_1^{(l)},\rr_2^{(l)},\ldots,{\rr_{N_l}^{(l)}})$. {More specifically,}
\beq
{e^{-\beta U_{ll}^\eff}=  \frac{e^{-\beta U_{ll}}}{N_s! \Lambda_s^{3N_s}} \int d\mathbf{r}^{N_{s}}e^{-\beta\left(U_{ss}+U_{sl}\right)}},
\label{general:eq4}
\eeq
so that, Eq.\ (\ref{general:eq2}) becomes
\begin{eqnarray}
e^{-\beta F}&=& \frac{1}{N_l! {\Lambda_l^{3N_l}}} \int d\mathbf{r}^{N_{l}} {e^{-\beta U_{ll}^\eff}}.
\label{general:eq5}
\end{eqnarray}
{In general, however, the effective potential $U_{ll}^\eff$ is not pairwise additive, meaning that apart from pair-interaction terms (and less relevant zero- and one-body terms), it requires three-body, four-body, \ldots terms. Thus, the general structure of $U_{ll}^\eff$ would be}
\beqa
{U_{ll}^\eff}&=&{N_l v_{ll}^{(0)}+\sum_{i=1}^{N_l}v_{ll}^{(1)}(\rr_i^{(l)})+ \sum_{i<j}^{N_l}v_{ll}^{(2)}(|\rr_i^{(l)}-\rr_j^{(l)}|)}\nn
&&{+ \sum_{i<j<k}^{N_l}v_{ll}^{(3)}(\rr_i^{(l)},\rr_j^{(l)},\rr_k^{(l)})+\cdots}.
\label{general:eq6}
\eeqa
{The physically most relevant contribution is expected to be the one associated with the effective \emph{pair} potential $v_{ll}(r)\equiv v_{ll}^{(2)}(r)$, in which case one can approximately neglect $v_{ll}^{(n)}$ with $n\geq 3$.}

{Now we specialize to the general class of NASHS models described in Sec.\ \ref{sec1}. The $\varphi_{ss}(r)$ and $\varphi_{ll}(r)$ pair interactions are of HS type characterized by diameters $\sigma_{ss}$ and $\sigma_{ll}$, respectively, while the small-large interaction $\varphi_{sl}(r)$ is of SHS type\cite{Baxter68,Baxter70} with a hard-core distance $\sigma_{sl}$ and a stickiness parameter $\tau_{sl}^{-1}$, the latter measuring the strength of surface adhesiveness. Therefore, the relevant Mayer functions $f_{\alpha\gamma}(r)=e^{-\beta\varphi_{\alpha\gamma}(r)}-1$  are}
\begin{eqnarray}
f_{ss}(r)&=&{-\Theta(\sigma_{ss}-r)},\\
f_{ll}(r)&=&{-\Theta(\sigma_{ll}-r)},\\
\label{general:eq3a}
f_{sl}(r)&=&{-\Theta(\sigma_{sl}-r)}+
\frac{\sigma_{sl}}{12\tau_{sl}}\delta(r-\sigma_{sl}).
\label{general:eq3b}
\end{eqnarray}
Here, {$\Theta(x)$  is the Heaviside step function and $\delta(x)$ is the Dirac delta function.
To simplify the notation, we adopt the viewpoint of the large spheres by calling $\sigma_{l}=\sigma_{ll}$ their diameter  and defining $\sigma_s$ as the diameter of the small spheres as felt by the large ones, so that $\sigma_{sl}=(\sigma_s+\sigma_l)/2$. Thus, the size asymmetry of the mixture (again from the viewpoint of the solute particles) is measured by the ratio $q=\sigma_s/\sigma_l<1$, while the nonadditivity of the hard-core interactions is measured by the ratio $\sigma_{ss}/\sigma_s\leq 1$ {(where, as said before, we discard here the case of negative nonadditivity)}. For later use, let us introduce the partial
packing fraction of species {$\alpha$} as} {$\eta_\alpha=\pi\rho_\alpha\sigma_\alpha^3/6$},
where {$x_\alpha=N_\alpha/N$} is the concentration of species {$\alpha=s,l$} and
{$\rho_\alpha=N_\alpha/V$} is its density. {The total number of
particles and number density of the fluid mixture are $N=N_l+N_s$ and $\rho=N/V$, respectively}.

{As discussed in Sec.\ \ref{sec1}, we now particularize to two interesting particular cases that are identified by the ratio $\sigma_{ss}/\sigma_{s}$: the ASHS model (where $\sigma_{ss}/\sigma_{s}=1$)
and the SAO model (where $\sigma_{ss}/\sigma_{s}=0$). The first model was studied before by two of us (it was called System A in Sec.\ V of Ref.\ \onlinecite{Fantoni05})
and has been rejuvenated by a recent study by Chen et al.\cite{Chen15} The second model is an extension of the well-known AO model, except that  a sticky (or adhesive)
interaction exists between the solvent and the solute particles. To the best of our knowledge, it has not been studied before. In both cases we will be able to derive the effective pair potential $v_{ll}(r)=v_{ll}^{(2)}(r)$ [see Eq.\ \eqref{general:eq6}] either within the PY approximation in the canonical ensemble (ASHS model) or in an exact way in the semi-grand-canonical ensemble (SAO model).}

\section{The PY approximate solution of the ASHS model}
\label{sec:ASHS}
The solution of the PY approximation for the ASHS model
was recently studied in Ref.\ \onlinecite{Fantoni05}. {The PY solution actually extends to the more general formulation where the}
Baxter {\emph stickiness} coefficient\cite{Baxter68,Baxter70} between
a particle of species $\alpha$ and one of species $\gamma$ is $\tau_{\alpha\gamma}^{-1}$.\cite{PS75,B75,Santos98}  Since here we choose
$\tau_{ss}\to\infty$ and $\tau_{ll}\to\infty$, we can only have
adhesion between unlike particles and $\tau_{sl}^{-1}>0$ measures its
strength.

\subsection{{Spinodal curve}}
\label{sec:PYcr}
\begin{figure}
\includegraphics[width=8cm]{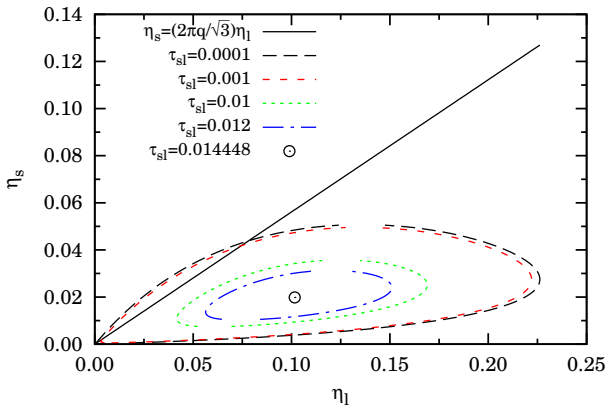}
\caption{PY spinodal for {$q=q_0$} and {several} values
  of $\tau_{sl}$. The straight line is $\eta_s=\eta_s^*$ and the circle
  is the critical point at {$\tau_{sl}=0.014\,448$}.
\label{fig:pyspinodal}}
\end{figure}

From Eq.\ (85) of Ref.\ \onlinecite{Fantoni05} we find the following
expression for the spinodal  of the full binary mixture in the $(\eta_s,\eta_l)$ plane, {as obtained from the PY
approximation,}
\beqa \label{pyspinodal}
{\tau_{sl}^\spi(\eta_s,\eta_l)}&=&{\Bigg[\frac{1+(1+q)(1-\eta_s-\eta_l)/3(\eta_s+q\eta_l)}
{1+\sqrt{\left(1+\frac{1-\eta_s-\eta_l}
{3\eta_s}\right)\left(1+\frac{1-\eta_s-\eta_l}{3\eta_l}\right)}}
-1\Bigg]}\nn
&&{\times
\frac{(1+q)(\eta_s+q\eta_l)}{4q(1-\eta_s-\eta_l)}},
\eeqa
which, as it should, is symmetric under the exchange {$\eta_s\leftrightarrow\eta_l$ and $q\leftrightarrow
1/q$}.  For a fixed $q$, there is a
maximum {value of $\tau_{sl}^\spi$} for which Eq.\ (\ref{pyspinodal}) admits a
solution with $\eta_s>0$ and $\eta_l>0$. We will denote this maximum
value with
{$\tilde{\tau}_{sl}$} and the {corresponding} solution, the critical
point, with
$(\tilde{\eta}_s,\tilde{\eta}_l)$. In particular, at {$q=q_0$ we find
$\tilde{\tau}_{sl}=  0.014\,448$,
$\tilde{\eta}_s=0.019\,839$, and
$\tilde{\eta}_l=0.101\,645$.} For
$\tau_{sl}<{\tilde{\tau}_{sl}}$ the solution of
Eq.\ (\ref{pyspinodal}) is a closed curve in the $(\eta_s,\eta_l)$
plane within which  the thermodynamically unstable region lies, as
shown in Fig.\ \ref{fig:pyspinodal}. As we can see, the spinodal {curve} does not change much for $\tau_{sl}<0.001$, {where it is crossed by the straight line representing the critical  packing
fraction} $\eta_s=\eta_s^*$.
These findings are in complete agreement with those reported in Ref.\ \onlinecite{Chen15}.

Note that Eq.\ (\ref{pyspinodal}) is a particular case of an equation for a general mixture derived
by Barboy and Tenne,\cite{Barboy79} that should however be handled with great care.\cite{Gazzillo04}

\subsection{Approximate effective one-component fluid}
\label{subsec:approximate}
{As explained in Sec.\ \ref{sec2}, one could in principle integrate out the solvent degrees of freedom to obtain the effective solute potential $U_{ll}^\eff$ {[see Eqs.\ \eqref{general:eq4} and \eqref{general:eq6}]}. Here we want to focus on the \emph{pair} interaction potential $v_{ll}(r)=v_{ll}^{(2)}(r)$. This function can be identified from the solute-solute radial distribution function $g_{ll}(r)$ in the infinite dilution limit ($x_l\to 0$) since in that limit only pair interactions contribute to  $g_{ll}(r)$. Therefore, $g_{ll}(r)\to e^{-\beta v_{ll}(r)}$ and hence}
\beq
{\beta v_{ll}(r)=-\lim_{x_l\to 0}\ln g_{ll}(r)}.
\label{vll}
\eeq
In the limit of no adhesion {($\tau_{sl}\to\infty$)}, $v_{ll}(r)$ becomes
the usual depletion potential.\cite{Yuste08,Fantoni14}
For further {use}, we will refer to
\emph{entropic} regime {as} the one  with $\tau_{sl}\gg 1$, close to a
size-asymmetric binary HS mixture. {Reciprocally,  the  \emph{nonentropic}
regime {will refer} to a system with a small $\tau_{sl}$. The \emph{transitional} regime {will correspond} to $\tau_{sl}\sim 1$}.

{Since {$\eta_s$ is supposed to be finite}  in Eq.\ \eqref{vll}, it is not possible to obtain the exact effective pair potential $v_{ll}(r)$. On the other hand, it can be obtained again from the PY solution, as described in Appendix \ref{app:RFA}. Note that, although the infinite dilution limit is applied as a short-cut to derive the pair potential $v_{ll}(r)$, {at a nonzero solute concentration} the full effective many-body potential $U_{ll}^\eff$ includes nonpairwise terms, as represented by $v_{ll}^{(3)}(\rr_i,\rr_j,\rr_k)$ and higher-order terms in Eq.\  \eqref{general:eq6}.}

\begin{figure}
\includegraphics[width=8cm]{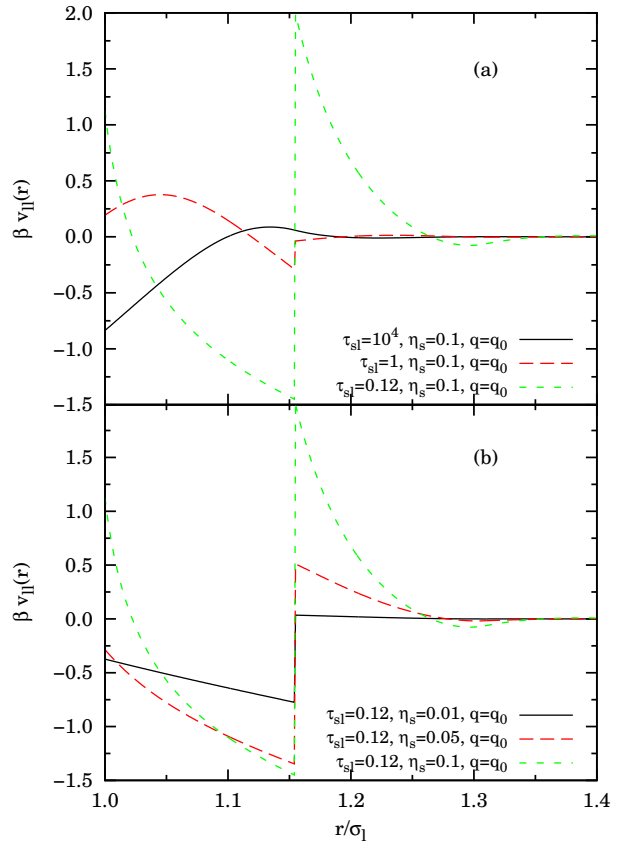}
\caption{Effective solute-solute pair potential {in the ASHS model, as obtained from the PY approximation (see Appendix \ref{app:RFA}). In panel (a) the stickiness parameter is varied at a fixed solvent  packing fraction $\eta_s=0.1$, while in panel (b), $\eta_s$ is varied at fixed $\tau_{sl}=0.12$. In all the cases the size ratio is {$q=q_0$}.} }
\label{fig:potential}
\end{figure}

In Fig.\ \ref{fig:potential} we report a few representative examples of the effective solute-solute
pair potential {corresponding to the ASHS model in the PY approximation (see Appendix \ref{app:RFA}). Figure  \ref{fig:potential}(a) shows the influence of the solute-solvent stickiness at fixed $\eta_s=0.1$ and {$q=q_0$}. One can clearly observe the different shape of the potential in the entropic ($\tau_{sl}=10^4$), transitional  ($\tau_{sl}=1$), and nonentropic ($\tau_{sl}=0.12$) regimes.  In the former case ($\tau_{sl}=10^4$), the potential is essentially attractive (except for a slight hump in the region $r/\sigma_l\lesssim 1+q$), thus reflecting the depletion mechanisms (see Fig.\ \ref{fig:fig1}). Moreover, at this very high value of $\tau_{sl}$, the discontinuity  of the potential at $r/\sigma_l=1+q$ [see Eq.\ \eqref{B15}] is not visible. In the transitional regime  ($\tau_{sl}=1$), however, the discontinuity at $r/\sigma_l=1+q$ is already noticeable and the potential in most of the inner region $1<r/\sigma_l<1+q$ has changed from attractive to repulsive. These two features are widely enhanced in the nonentropic regime ($\tau_{sl}=0.12$): there is a high discontinuity at $r/\sigma_l=1+q$ and the effective potential is strongly repulsive in the whole region $1<r/\sigma_l<1+q$. Furthermore, a strong repulsion appears as well in the outer region $r/\sigma_l\gtrsim 1+q$. Figure \ref{fig:potential}(b) shows that an increase of the solvent density magnifies the characteristic features of the effective potential in the nonentropic regime. The physical origin of the repulsive regions in the nonentropic regime can be ascribed to the net \emph{pulling} role played by the solvent particles attached to the two solutes. This effect will be identified more clearly in the SAO model (see Sec.\ \ref{sec:AAO}). As for the (attractive) discontinuity at $r/\sigma_l= 1+q$, it can be attributed to the \emph{bridging}  effect of solvent particles attached to both solutes. This bridging mechanism is absent if $r/\sigma_l= (1+q)^+$ but appears if  $r/\sigma_l= (1+q)^-$.}

Dijkstra et al.\cite{Dijkstra99a,Dijkstra99} already showed
that the effective potential in the entropic regime is unable to produce a stable
demixing phase transition with reasonably small
$q$. On the other hand, the {step attraction at $r/\sigma_l= 1+q$ in the} potential associated with the nonentropic regime can lead to  a demixing transition, as shown in
Ref.\ \onlinecite{Fantoni05}. This is the phase
instability studied in the $(\eta_s,\eta_l)$ plane in Sec.\
\ref{sec:PYcr}.

With all due cares, the shape of the effective potential
in the nonentropic regime depicted in {Fig.\ \ref{fig:potential}(b) can be} \emph{schematically} represented as a square-well (SW) potential {of width $q\sigma_l$ and depth $\epsilon\sim |v_{ll}(\sigma_l(1+q)^-)|$},
with an additional  repulsive tail {starting at} $r=\sigma_l(1+q)^+$. {We can then exploit the fact that the} phase behavior of a one-component SW fluid is well established.\cite{Vega92,Liu05,Acedo01,Espindola09}
For example, it is sufficient to heuristically consider the approximate critical value\cite{Acedo01} of the reduced temperature $T^*=k_BT/\epsilon$ to find {the appearance of an open phase
coexistence region at high $\eta_s$ (well separated from the closed one predicted
in Sec.\ \ref{sec:PYcr} at low $\eta_s$). This coexistence region is known to be present in the
highly asymmetric AHS mixture\cite{Dijkstra99,Dijkstra99a} (i.e., for small $q$ in the limit $\tau_{sl}\to\infty$).
The effective problem procedure that we followed suggests that, quite intuitively, such a
region will not disappear when the attraction is switched on at small $\tau_{sl}$.} It is
interesting to observe that  such a {reentrance} at large $\eta_s$ is
not predicted by an analysis of the behavior of the effective {second virial coefficient $B_2^\eff$ [see Eq.\ \eqref{B16}], according to which $1/T^*=\ln\left[1+(1-B_2^\eff/B_2^\HS)/(3q+3q^2+q^3)\right]$,  where $B_2^\HS=\frac{2\pi}{3}\sigma_l^3$.} The two heuristic criteria {based on an effective SW temperature $T^*$} agree quite well for small
values of $\eta_s$ {[as expected from the curve $\eta_s=0.01$ in Fig.\ \ref{fig:potential}(b)],} but the $B_2^\eff$ criterion presents a diverging
{$T^*$ at a value of $\eta_s$ such that $B_2^\eff=B_2^\HS$} and becomes meaningless thereafter {(i.e., when $B_2^\eff>B_2^\HS$). For instance, if {$q=q_0$ and $\tau_{sl}=0.12$, the condition $B_2^\eff>B_2^\HS$ is satisfied for {$\eta_s>0.274$}. The fact that $B_2^\eff>B_2^\HS$ if $\eta_s$ is large enough is directly related to the increase of the effective size of the \emph{dressed} solute particles, as depicted in Figs.\ \ref{fig:fig2}(c) and \ref{fig:fig2}(d)}.

Of course, {the} effective one-component fluid is not {fully} equivalent to the
original binary mixture, as we are neglecting three-body (and higher) terms in
the effective total potential {[see Eq.\ \eqref{general:eq6}]}. Moreover, the potentials of
Fig.\ \ref{fig:potential} are the outcome of the PY approximation. Yet,
they are expected to give reasonable approximate results in
the spirit of an effective fluid.
Chen et al.\cite{Chen15} devised a similar approximate mapping of the
{PY}  solution {for the true binary mixture onto} a one-component {SHS} model, from which
they were able to read-off the binodal using accurate Monte Carlo {(MC)} results
by Miller and Frenkel.\cite{Miller04}

{While some caution must be exercised when  using the pairwise potential formally obtained in the limit $\eta_l\to 0$ to predict the phase diagram at finite $\eta_l$, this  keeps being a useful procedure to reduce the complexity of the binary mixture problem,\cite{SVB15} allowing one
to get additional physical insight without the need, for example, of performing computer simulations of
the full binary mixture.}

\section{The SAO model }
\label{sec:AAO}
{As shown in Sec.\ \ref{sec:ASHS}, the} ASHS model {($\sigma_{ss}/\sigma_s=1$)} admits a PY analytical solution but only an approximate
reduction to an effective one-component fluid.
The SAO model {($\sigma_{ss}/\sigma_s=0$)} is, in some sense, complementary to it, as it does not admit an
analytical solution, not even in the PY approximation, but it does admit
an exact reduction to an effective one-component fluid for
$q<q_0=2/\sqrt{3}-1\simeq 0.1547$, when a solvent {particle} can fit into the
inner volume created by three solutes at contact,\cite{Gast83} {so that a solvent particle cannot overlap simultaneously with more than two (nonoverlapping) solute particles.  This corresponds} to $q<1$ in one
spatial dimension.\cite{Brader02}

{To proceed, it is convenient to change from the canonical $(N_s,N_l,V,T)$ ensemble to the semi-grand-canonical $(\mu_s,N_l,V,T)$ ensemble,\cite{Dijkstra99,Dijkstra99b} where $\mu_s$ is the chemical potential of the solvent component. The corresponding thermodynamic potential ${\cal F}(\mu_s,N_l,V,T)$ is constructed  via the Legendre transform}
\begin{eqnarray}
{\cal F}(\mu_s,N_l,V,T)&=& F\left({\langle N_s\rangle},N_l ,V,T\right)-\mu_s {\langle N_s\rangle}.
\label{general:eq7}
\end{eqnarray}
{Thus, the counterpart of  canonical Eq.\ \eqref{general:eq2} is}
\begin{eqnarray}
\label{general:eq10}
e^{-\beta {\cal F}} &=& {\sum_{N_s=0}^{\infty} \frac{z_s^{N_s}}{N_s! N_l!\Lambda_l^{3N_l}}\int d\rr^{N_s}\int d\rr^{N_l}e^{-\beta (U_{ll}+U_{sl})}}\nn
&=&{\frac{1}{N_l! \Lambda_l^{3N_l}} \int d\mathbf{r}^{N_{l}} e^{-\beta U_{ll}^\eff}},
\end{eqnarray}
where
\begin{eqnarray}
z_s&=& \frac{e^{\beta \mu_s}}{\Lambda_s^3}
\label{general:eq8}
\end{eqnarray}
is the solvent fugacity and
\beq
\label{general:eq9}
e^{-\beta {U_{ll}^\eff}} = {e^{-\beta U_{ll}}}\sum_{N_s=0}^{\infty} \frac{{z_s^{N_s}}}{N_s!} \int d\mathbf{r}^{N_{s}} e^{-\beta U_{sl}}.
\eeq
{Note that in Eq.\ \eqref{general:eq10} we have taken into account that $U_{ss}=0$ in the SAO model.}

{Inserting Eq.\ \eqref{general:eq1} into Eq.\ \eqref{general:eq9} it is easy to obtain\cite{Dijkstra99b}}
\beq
\label{general:eq12}
\beta \Omega =- z_s \int d \mathbf{r}\, \prod_{i=1}^{N_l}\left[1+f_{sl} ( |\rr-\rr_i^{(l)}| )  \right],
\eeq
{where $\Omega=U_{ll}^\eff-U_{ll}$ represents the grand potential of an ideal gas of solvent particles  in the
external field of a fixed configuration of $N_l$ solute particles with coordinates $\{\rr_i^{(l)}\}$. Expanding in products of Mayer functions, $\Omega$ can be written as}
\beq
\label{general:eq12.b}
\Omega =
\sum_{n=0}^{{n_{\max}(q)}} \Omega_{n}.
\eeq
Here, $\Omega_{n}$ is the contribution to $\Omega$ stemming from  {the product of $n$ Mayer functions $f_{sl}$.
The upper limit $n_{\max}(q)$  is the maximum number of nonoverlapping solutes that can simultaneously overlap with a single solvent particle. For $n>n_{\max}(q)$, at least one of the factors $f_{sl}$ vanishes and so does $\Omega_n$. If $q<q_0$, then $n_{\max}(q)=2$, implying that the exact effective potential $U_{ll}^\eff$ does not include three-body (or higher order) terms. In the interval $q_0<q\leq 1$, $n_{\max}(q)$ grows by steps as $q$ increases,
reaching a maximum value $n_{\max}(q)=11$ (since a solvent particle can simultaneously overlap with 12 nonoverlapping solutes only if $q>1$).}
{The first few terms in Eq.\ \eqref{general:eq12.b} are}
 \bq
\beta\Omega_0&=&-z_sV,\label{4.7}\\
\beta\Omega_1&=&- z_s\sum_{i=1}^{N_l} \int d \mathbf{r} f_{sl}(|\mathbf{r}-\mathbf{r}_i^{(l)}|),\\
\beta\Omega_2
&=&- z_s\sum_{i<j}^{N_l} \int d \mathbf{r} f_{sl}(|\mathbf{r}-\mathbf{r}_i^{(l)}|) f_{sl}{(}|\mathbf{r}-\mathbf{r}_j^{(l)}|{)}\nn
& {=}&{\beta\sum_{i<j}^{N_l}\left[v_{ll}(|\rr_i^{(l)}-\rr_j^{(l)}|)-\varphi_{ll}(|\rr_i^{(l)}-\rr_j^{(l)}|)\right]}.
\label{Omega2}
\eq
{Equation \eqref{Omega2} allows us to identify the exact effective pair potential as}
\beq
{\beta v_{ll}(r)=\beta \varphi_{ll}(r)-z_s\int d\rr_s\,f_{sl}(r_s)f_{sl}(|\rr_s-\rr|)}.
\label{vllff}
\eeq

\begin{figure}
\includegraphics[width=8cm]{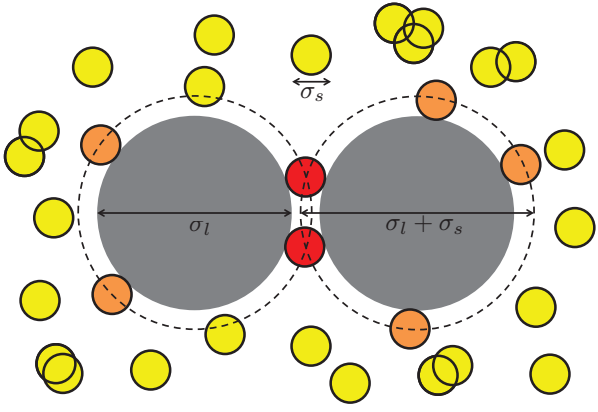}
\caption{Cartoon describing the three effects {(depletion, pulling, and bridging) contributing to the effective solute-solute interaction in the SAO model}. The grey large spheres represent
the solutes of diameter $\sigma_l$ at a distance
$r<\sigma_l+\sigma_s{=\sigma_l(1+q)}$. {They} are surrounded by {a sea of} smaller
  spheres (the solvent) of diameter $\sigma_s=q\sigma_l$ ({$q=0.2$} in the
  cartoon) that can overlap among themselves and have a sticky
  surface (represented  by a thick perimeter). Some of
  the solvent {particles} (the yellow ones) do
  not touch the solutes and so they contribute to
  the (attractive) depletion effect, which is {represented by $\psi_{\text{d}}(r)$,} a volumetric term
  independent of $\tau_{sl}$. Other {solvent particles} (the orange ones) are
  adhered to one of the big spheres, {thus
  contributing} to the (repulsive) pulling effect, {represented by $\psi_{\text{p}}(r)$}, which is a surface
  {term} proportional to $\tau_{sl}^{-1}$. Finally, some other small
  particles (the red ones) are adhered to both
  solutes, giving rise to the (attractive) bridging effect, {represented by $\psi_{\text{b}}(r)$}, which is a
  {line} {(intersection of two surfaces)} term proportional to $\tau_{sl}^{-2}$.}
\label{fig:aao}
\end{figure}

\begin{figure*}
\includegraphics[width=8cm]{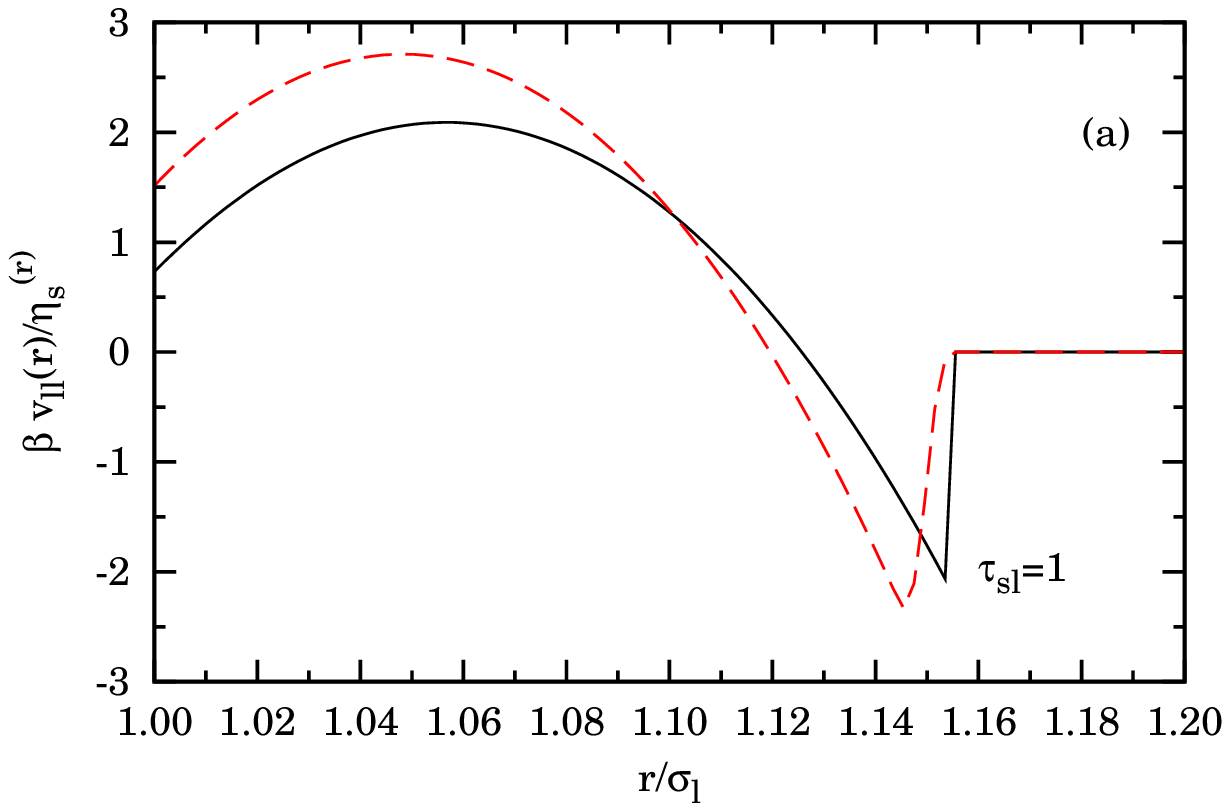}\hfill
\includegraphics[width=8cm]{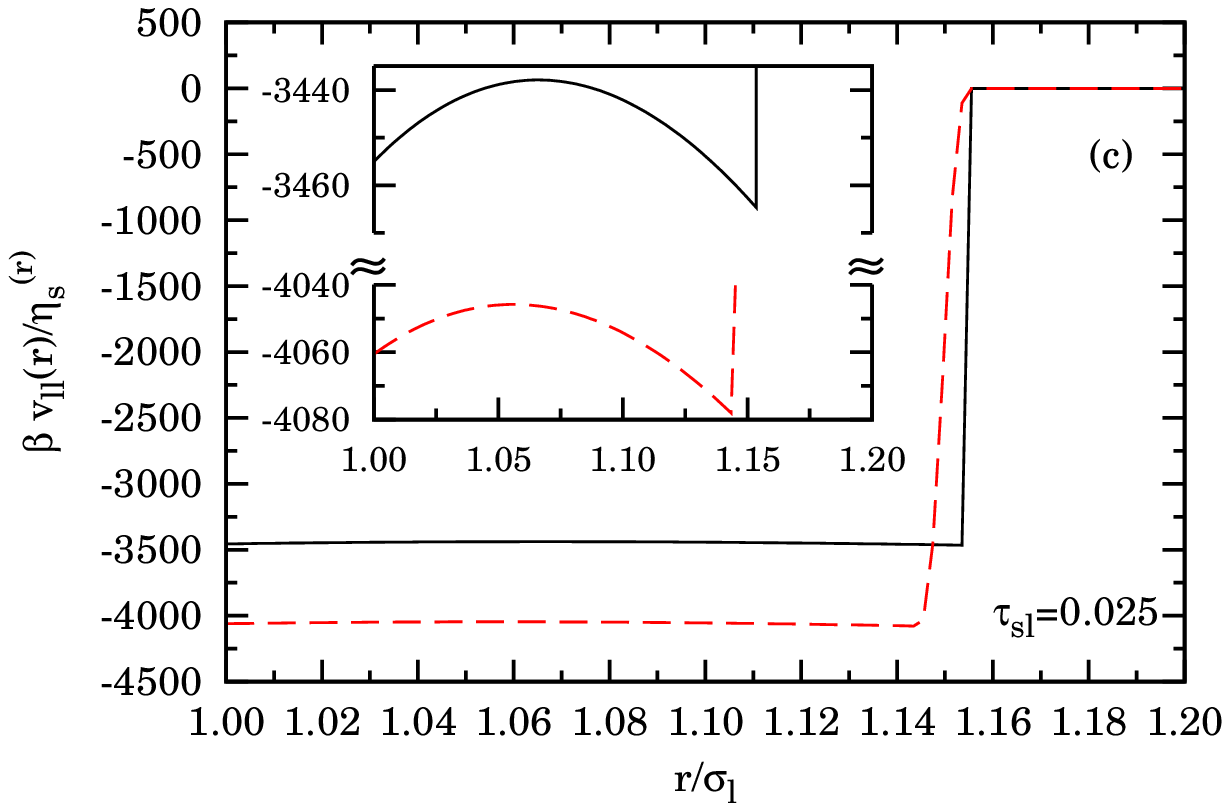}\\
\includegraphics[width=8cm]{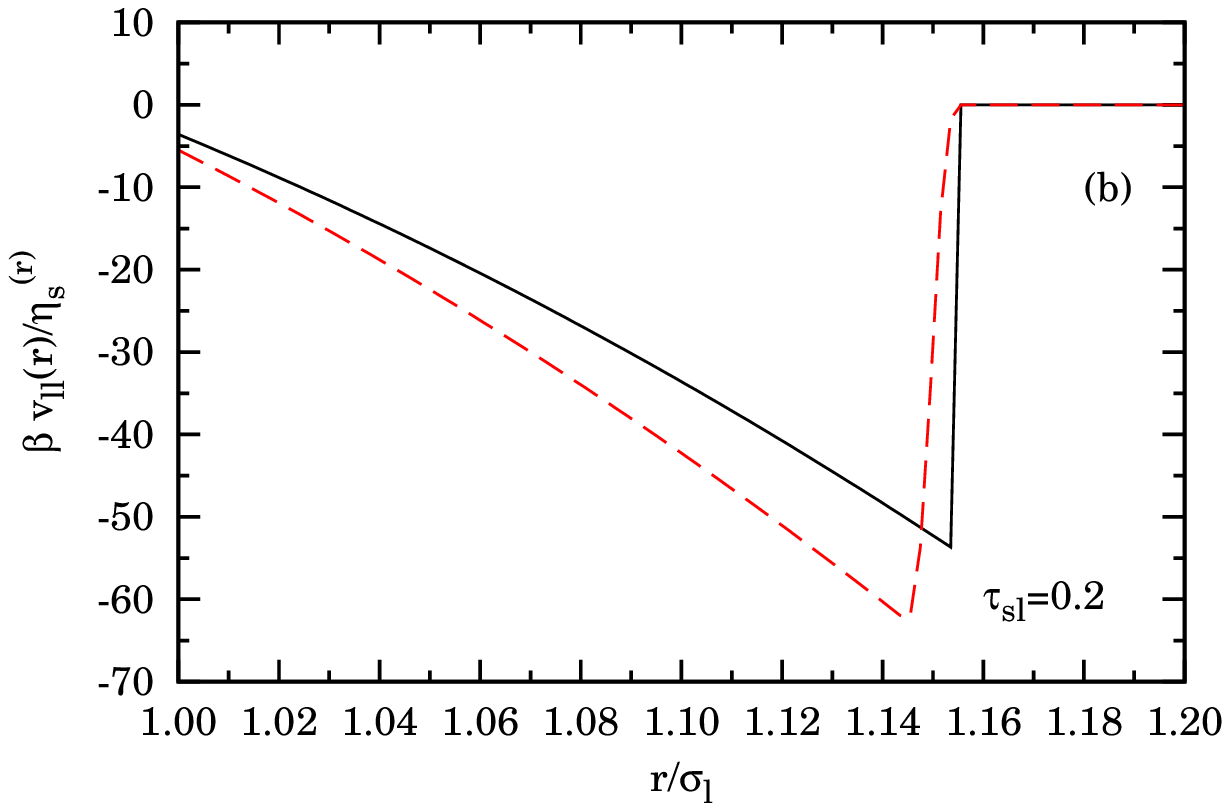}\hfill
\includegraphics[width=8cm]{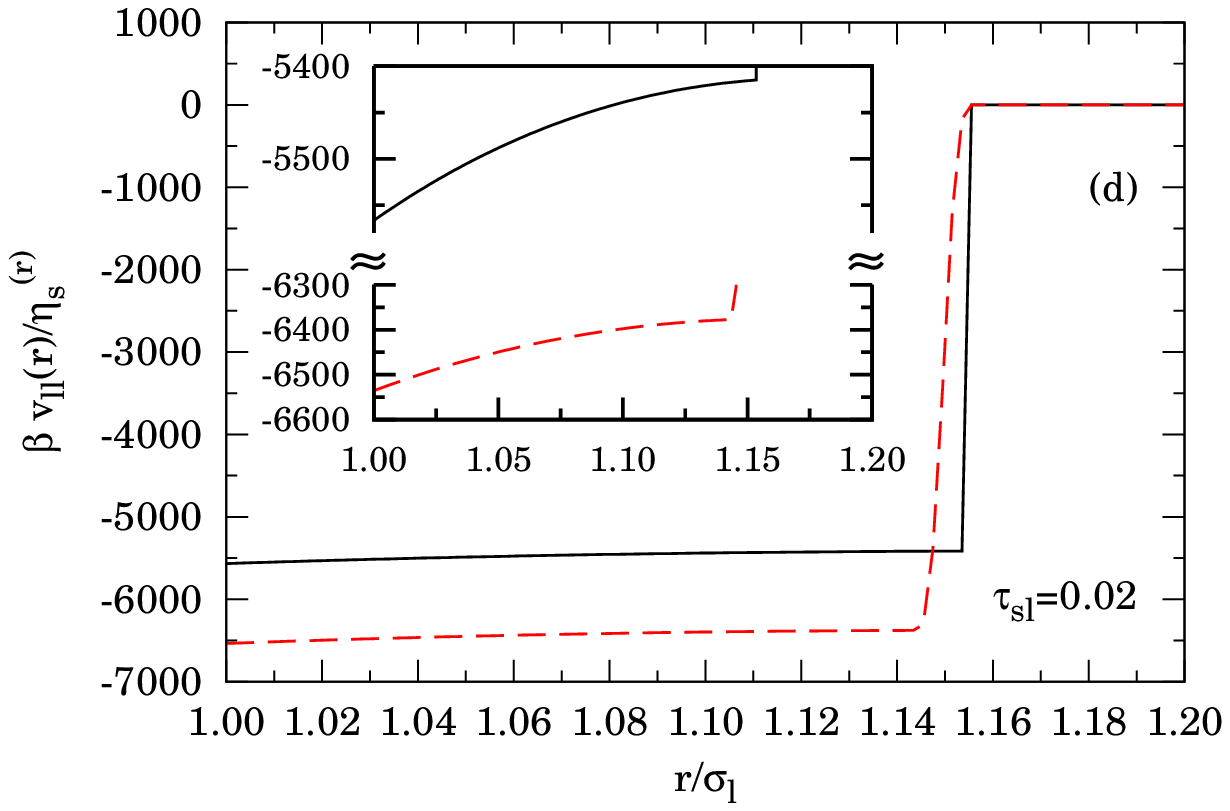}
\caption{Plot of $\psi(r)\equiv \beta v_{ll}(r)/\eta_s^{(r)}$  for (a) $\tau_{sl}=1$, (b) $\tau_{sl}=0.2$, (c) $\tau_{sl}=0.025$, and (d) $\tau_{sl}=0.02$. {The solid lines correspond to the SAO model at the threshold value  $q=q_0$, while the dashed lines correspond to the SWAO model with $q=q_0-\xi$, $\Delta_{sl}=\xi/(1+q)$, $\xi=10^{-2}$. The insets in panels (c) and (d) show magnified views of the curves for $r/\sigma_l<1+q_0$.} }.
\label{fig:potaao}
\end{figure*}

\begin{figure}
\includegraphics[width=8cm]{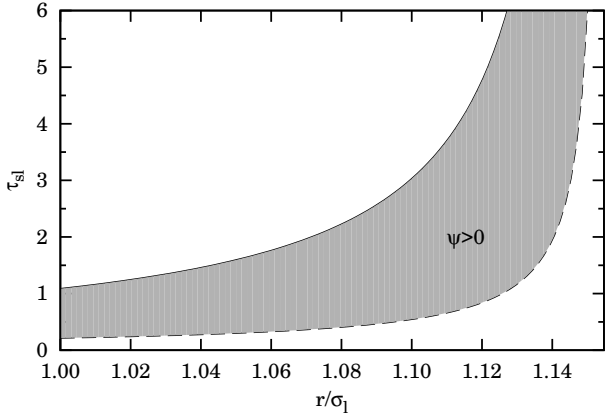}
\caption{{Plane $\tau_{sl}$ vs $r$ showing the region where the effective pair potential in the SAO model for the threshold value $q=q_0$ takes positive values. Outside the shaded region the potential is negative.}}
\label{fig:sign}
\end{figure}

{Now, making use of Eq.\  (\ref{general:eq3b}), one can obtain}
\beq
{\beta \Omega_1=z_s\eta_l V(1+q)^3\left(1-\frac{1}{4\tau_{sl}}\right)},
\label{4.11}
\eeq
\beq
\beta v_{ll}(r)=\eta_s^{(r)}\begin{cases}
  \infty,&r<\sigma_l,\\
    {\psi(r)},&\sigma_l<r<\sigma_l(1+q),\\
  0,&r>\sigma_l(1+q),
\end{cases}
\label{epaao}
\eeq
{where $\eta_s^{(r)}=z_s (\pi/6)\sigma_s^3$ is the (nominal) solvent  packing fraction of a \emph{reservoir} made of noninteracting solvent particles and}
\beq
{\psi(r)=\psi_{\text{d}}(r)+\psi_{\text{p}}(r)+\psi_{\text{b}}(r)}
\label{psidpb}
\eeq
with
\beqa
\psi_{\text{d}}(r)&=&-\frac{6}{\pi\sigma_s^3}\int d\rr_s\,\Theta(\sigma_{sl}-r_s)\Theta(\sigma_{sl}-|\rr_s-\rr|)\nn
&=&-\frac{(1+q-r/\sigma_l)^2(2+2q+r/\sigma_l)}{2q^3},\label{psid}\\
\psi_{\text{p}}(r)&=&\frac{\sigma_{sl}}{\pi\sigma_s^3\tau_{sl}}\int d\rr_s\,\delta(r_s-\sigma_{sl})\Theta(\sigma_{sl}-|\rr_s-\rr|)\nn
&=&\frac{(1+q)^2(1+q-r/\sigma_l)}{4q^3\tau_{sl}},\label{psip}\\
\psi_{\text{b}}(r)&=&-\frac{\sigma_{sl}^2}{24\pi\sigma_s^3\tau_{sl}^2}\int d\rr_s\,\delta(r_s-\sigma_{sl})\delta(|\rr_s-\rr|-\sigma_{sl})\nn
&=&-\frac{(1+q)^4}{192q^3\tau_{sl}^2r/\sigma_l}.\label{psib}
\eeqa
{The effective solute-solute force $f_{ll}(r)=-\partial v_{ll}(r)/\partial r$ (outside the hard core, $r>\sigma_l$) is }
\beqa
\frac{\beta f_{ll}(r)}{\eta_s^{(r)}}&=&-\left[\psi'_{\text{d}}(r)+\psi'_{\text{p}}(r)+\psi'_{\text{b}}(r)\right]\Theta(1+q-r/\sigma_l)\nn
&&-
\frac{(1+q)^3}{192q^3\tau_{sl}^2}\delta(r-\sigma_l(1+q)),
\eeqa
{where the delta term reflects the discontinuity of $v_{ll}(r)$ at $r=\sigma_l(1+q)$ and}
\beq
\psi'_{\text{d}}(r)=\frac{3}{2q^3\sigma_l}\left[(1+q)^2-\frac{r^2}{\sigma_l^2}\right],
\label{psid'}
\eeq
\beq
\psi'_{\text{p}}(r)=-\frac{(1+q)^2}{4q^3\sigma_l\tau_{sl}},\quad
\psi'_{\text{p}}(r)=\frac{(1+q)^4}{192q^3\tau_{sl}^2r^2/\sigma_l},
\label{psipb'}
\eeq

{If $q<q_0$,} the {general} relationship between {the reservoir packing fraction} $\eta_s^{(r)}$ {(or, equivalently, the fugacity $z_s$)} and the {values} $\eta_s$
{and $\eta_l$ of the binary mixture is derived in Appendix \ref{app:alternative} with the result}
\beqa
\label{resden}
\eta_s&=&\eta_s^{(r)}\Bigg[1-
\eta_l(1+q)^3\left(1-\frac{1}{4\tau_{sl}}\right)-\frac{12\eta_l^2 q^3}{\sigma_l^3}\nn
&&\times\int_{\sigma_l}^{\sigma_l(1+q)}dr\, r^2 \psi(r)g_{\eff}(r|\eta_l,\eta_s^{(r)})\Bigg],
\eeqa
{where $g_{\eff}(r|\eta_l,\eta_s^{(r)})$ is the radial distribution function of a pure fluid of large particles interacting via the effective pair potential $v_{ll}(r)$ at a packing fraction $\eta_l$. Up to second order in $\eta_l$, {Eq.\ \eqref{resden} becomes}}
\beqa
\label{resden2}
\eta_s&\approx&\eta_s^{(r)}\Bigg[1-
\eta_l(1+q)^3\left(1-\frac{1}{4\tau_{sl}}\right)-\frac{12\eta_l^2 q^3}{\sigma_l^3}\nn
&&\times\int_{\sigma_l}^{\sigma_l(1+q)}dr\, r^2 \psi(r)e^{-\eta_s^{(r)}\psi(r)}\Bigg],
\eeqa

Interestingly,  exact effective pair-potential
(\ref{epaao}) can be equivalently obtained from a density expansion of the
approximate PY effective potential of the ASHS model described in Sec.\ \ref{sec:ASHS}, upon
neglecting terms of order higher than linear in
$\eta_s$ and identifying $\eta_s$ with $\eta_s^{(r)}$, that is correct in
the solute infinite dilution limit {[see Eq.\ \eqref{resden}]}. This is not a coincidence\cite{Yuste08} {because the PY theory}
gives the exact radial distribution function to first order in density
(and therefore it gives the exact effective potential to that
order) {and} the relevant Mayer diagram, { containing only one solvent particle, is the same} whether the mixture is additive or not.

The {three} terms appearing in Eq.\ (\ref{psidpb})
bear a particularly simple and instructive physical interpretation.
The first term, {$\psi_{\text{d}}(r),$ [see Eqs.\  \eqref{psid} and \eqref{psid'}]} is the conventional
AO effective potential.\cite{Dijkstra99}
If $r<\sigma_l(1+q)$, no solvent particles fit
in the line joining the centers of the two solute particles. This is
the typical configuration of \emph{depletion} when the solute-solvent
interactions are of HS type, giving rise to an effective
\emph{attraction} between the solutes {(with a force decreasing its strength quadratically with increasing distance)}. Now imagine {we} switch the stickiness on. {Interestingly,} this
 produces two competing effects. Firstly,
the solvent particles attached to the outer surfaces
of each facing solute {tend} to \emph{pull} the solutes apart, producing an effective
solute-solute \emph{repulsion} {with a constant force strength}. {This is represented by the ``pulling'' term $\psi_{\text{p}}(r)$ [see Eqs.\  \eqref{psip} and \eqref{psipb'}]}.
Secondly, the solvent particles
attached to \emph{both} facing solutes (the ``bridges'') tend to increase
\emph{attraction} {(with a Coulomb-like force strength decreasing with increasing distance)}, {this bridging effect being represented by the term $\psi_{\text{b}}(r)$ [see Eqs.\  \eqref{psib} and \eqref{psipb'}]}.
{These three effects are schematically synthesized} in Fig.\ \ref{fig:aao}.

{It is interesting to remark that the SAO model can be easily extended by replacing the solute-solvent sticky surface by a finite-width ($\Delta_{sl}$) SW interaction. The resulting SWAO model is worked out in Appendix \ref{app:SWAO}. In this case, the condition for an exact reduction of the effective solute interaction to pairwise terms is $q(1+\Delta_{sl})+\Delta_{sl}<q_0$.}

The interplay of the three contributions to $\psi(r)$ gives rise to interesting transitions in the shape of the depletion potential, as illustrated in Fig.\ \ref{fig:potaao} {for the SAO and SWAO models. Let us  comment the curves corresponding to the SAO model.}
 {For relatively weak
stickiness, $\tau_{sl}^{-1}<\tau_{-}^{-1}\equiv 24(1-\sqrt{1-q-q^2/2})/(1+q)^2$, the pulling effect  dominates over the bridging effect for all distances but is dominated by the depletion  effect, except for distances close to $r=\sigma_l(1+q)$. Consequently, the effective potential is attractive near $r=\sigma_l$ and repulsive near $r=\sigma_l(1+q)$, as happens in Fig.\ \ref{fig:potaao}(a). Next, in the intermediate regime $\tau_{-}^{-1}<\tau_{sl}^{-1}<\tau_{+}^{-1}\equiv 24(1+\sqrt{1-q-q^2/2})/(1+q)^2$ the pulling effect dominates for all distances and the potential is purely repulsive, except for the discontinuous jump at $r=\sigma_l(1+q)$. This is represented by the case of Fig.\ \ref{fig:potaao}(b). In the strong stickiness regime $\tau_{+}^{-1}<\tau_{sl}^{-1}<48$ the depletion effect is practically irrelevant and the pulling effect is dominated by the bridging one, except in the region $r\lesssim\sigma_l(1+q)$. As a consequence, the effective potential is \emph{slightly} attractive near $r=\sigma_l$ and \emph{slightly} repulsive near $r=\sigma_l(1+q)$, as happens in Fig.\ \ref{fig:potaao}(c). Finally, for very strong stickiness ($\tau_{sl}^{-1}>48$) the bridging dominates over the pulling for all distances and the potential is purely attractive. This is the case displayed in Fig.\ \ref{fig:potaao}(d). {Those features are essentially preserved in the case of the SWAO model, except that the jump at $r=\sigma_l(1+q)$  is replaced by a rapid (but continuous) increase of the potential between  $r=\sigma_l(1+q)$ and $r=\sigma_l(1+q)(1+\Delta_{sl})$.}

{From Eqs.\ \eqref{psid}--\eqref{psipb'} it is easy to see that in the SAO model the effective potential and force are positive if}
\beq
{\frac{6\hat{r}-\sqrt{6\hat{r}(5\hat{r}-2)}}{24\hat{r}(1-\hat{r})(2+\hat{r})}\leq \tau_{sl}\leq
\frac{6\hat{r}+\sqrt{6\hat{r}(5\hat{r}-2)}}{24\hat{r}(1-\hat{r})(2+\hat{r})}}
\eeq
{and}
\beq
{\frac{2\hat{r}-\sqrt{2(3\hat{r}^2-1)}}{24\hat{r}{(1-\hat{r}^2)}}\leq \tau_{sl}\leq \frac{2\hat{r}+\sqrt{2(3\hat{r}^2-1)}}{24\hat{r}{(1-\hat{r}^2)}}},
\eeq
{respectively, where $\hat{r}\equiv r/[\sigma_l(1+q)]$. Figure \ref{fig:sign} shows the region in the plane $\tau_{sl}$ vs $r$ where $\psi(r)>0$ for the threshold value $q=q_0$.}

{As can be seen from Figs.\ \ref{fig:potaao}(c) and \ref{fig:potaao}(d), the effective potential in the regime of strong stickiness clearly resembles that of a SW potential of width $q\sigma_l$ and depth $\beta\epsilon=\eta_s^{(r)}|\psi(\sigma_l(1+q))|=\eta_s^{(r)}(1+q^{-1})^3/192 \tau_{sl}^2$}.  In this case, the effective phase behavior of an
equivalent SW fluid would suggest that, for a given $q<q_0$ and
a sufficiently small $\tau_{sl}$, we have the appearance of just one
lower critical point $(\eta_s^{(r)c},\eta_l^c)$ and
the instability region does not close itself again at
$\eta_s^{(r)}>\eta_s^{(r)c}$. This is a scenario quite different from the one in the model ASHS, where we found at least one closed island with a lower
and an upper critical point (see Sec.\ \ref{sec:PYcr}). It would then be
sufficient to switch on a hard-core repulsion {(with $\sigma_{ss}=\sigma_s$)} {among} the solvent
particles to have a closed spinodal.
{Along similar lines, it} is also interesting to observe that {the threshold packing fraction $\eta_s^*$ defined in Sec.\ \ref{sec1} clearly diverges in the SAO model because the solvent particles can freely overlap.}

\section{The Noro--Frenkel criticality criterion}
\label{sec:NF}
In 2000, Noro and Frenkel (NF)\cite{Noro00} argued that the reduced second virial coefficient
{$B_2/B_2^\HS$}, rather than the range and the strength of the attractive interactions, could be the most convenient quantity
to estimate the location of the critical point for a {wealth} of different colloidal suspensions.
{Their criticality criterion for particles with
variable range attractions,\cite{Noro00} complemented by the simulation value of the critical temperature obtained in Ref.\ \onlinecite{Miller04} for the SHS model, yields $B_2/B_2^\HS\simeq -1.21$.}

{In this section we apply {the} NF criterion to the two models discussed before: the ASHS model (see Sec.\ \ref{sec:ASHS}) and the SAO model (see Sec.\ \ref{sec:AAO}). In both cases, if $v_{ll}(r)$ is the effective solute-solute pair potential, the associated second virial coefficient $B_2^\text{eff}$ is given by}
\bq
\frac{B_2^\text{eff}}{B_2^\HS}=1-\frac{3}{\sigma_l^3}\int_{\sigma_l}^\infty dr\,r^2
\left[e^{-\beta v_{ll}(r)}-1\right],
\eq
where $B_2^\HS=2\pi\sigma_l^3/3$ is the virial coefficient for HSs of
diameter $\sigma_l$.
{Paradoxically, while the explicit PY expression of $\beta v_{ll}(r)$ in the ASHS model is rather cumbersome (see Appendix \ref{app:RFA}), its associated second virial coefficient $B_2^\text{eff}$ is much easier to obtain thanks to properties of the Laplace representation. The result can be found in Eq.\ \eqref{B16}. In contrast, in the SAO model the exact expression of $\beta v_{ll}(r)$ is very simple [see Eqs.\ \eqref{epaao}--\eqref{psib}] but the computation of $B_2^\text{eff}$ needs to be done numerically.}

\begin{figure*}
\includegraphics[width=8cm]{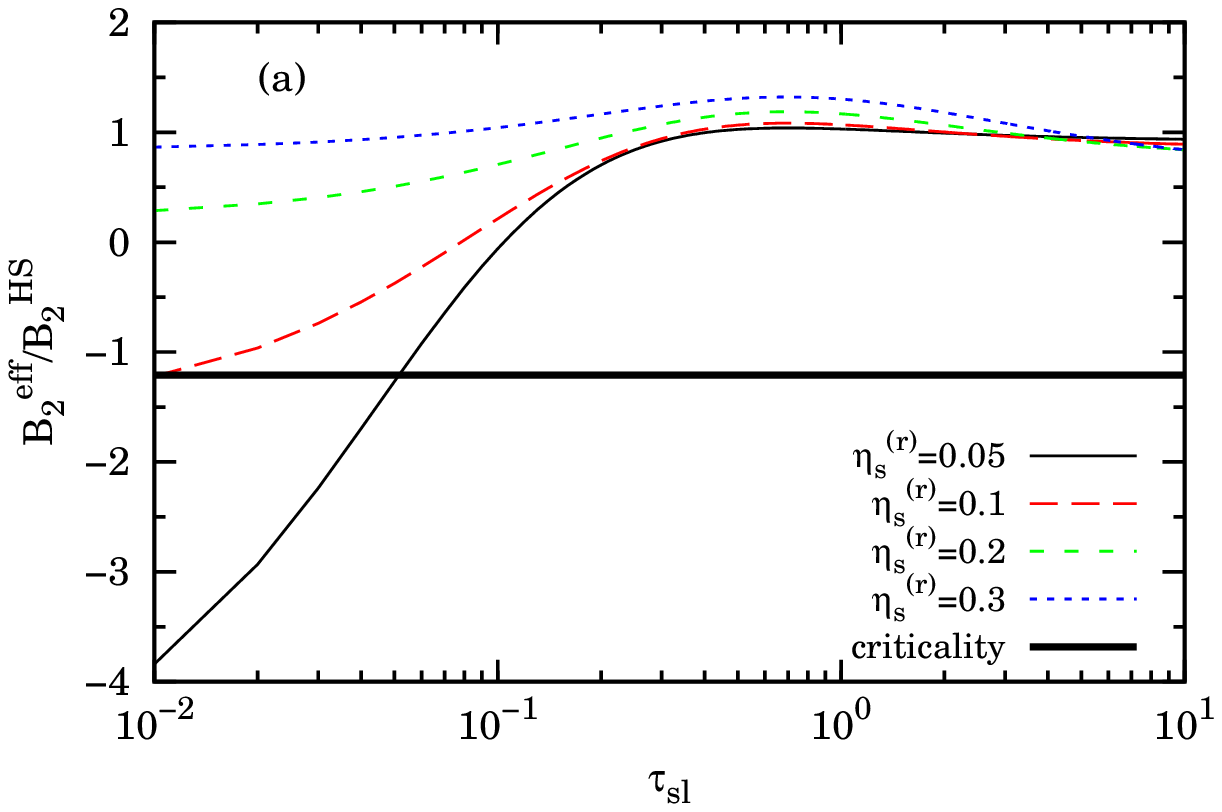}\hfill\includegraphics[width=8cm]{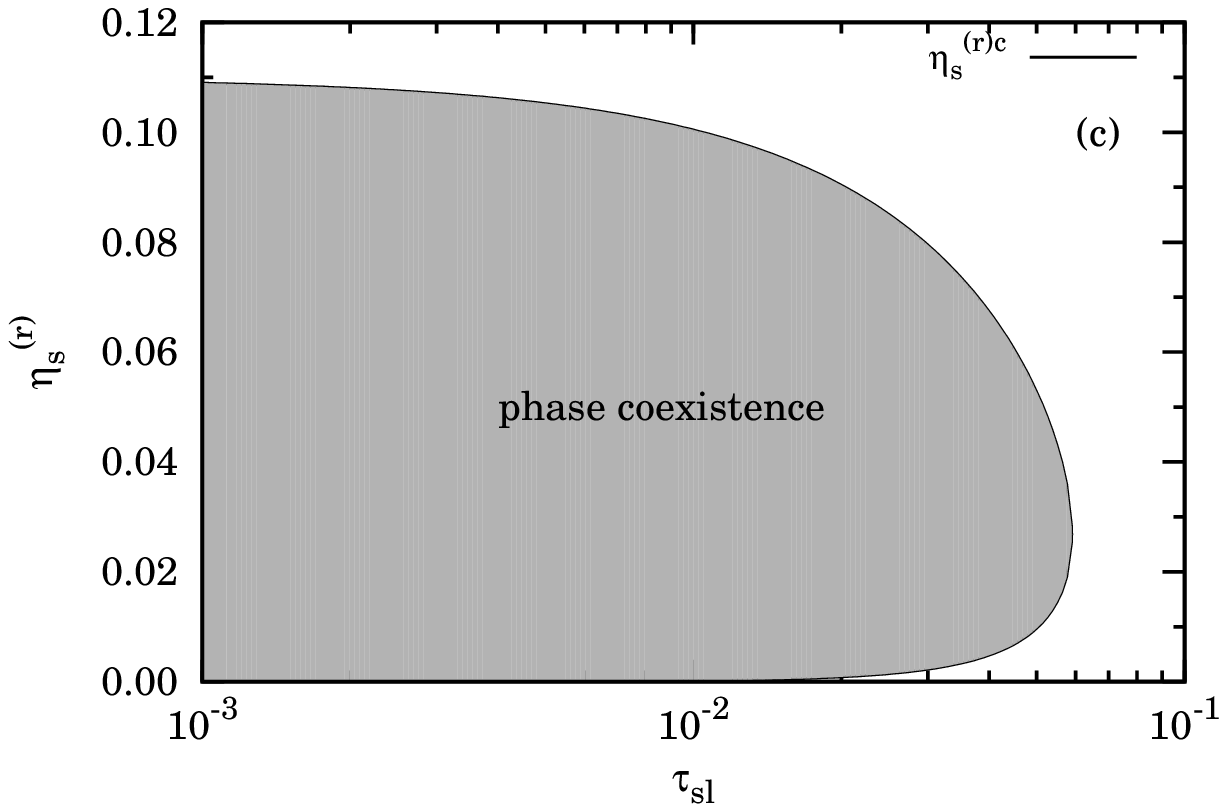}\\
\includegraphics[width=8cm]{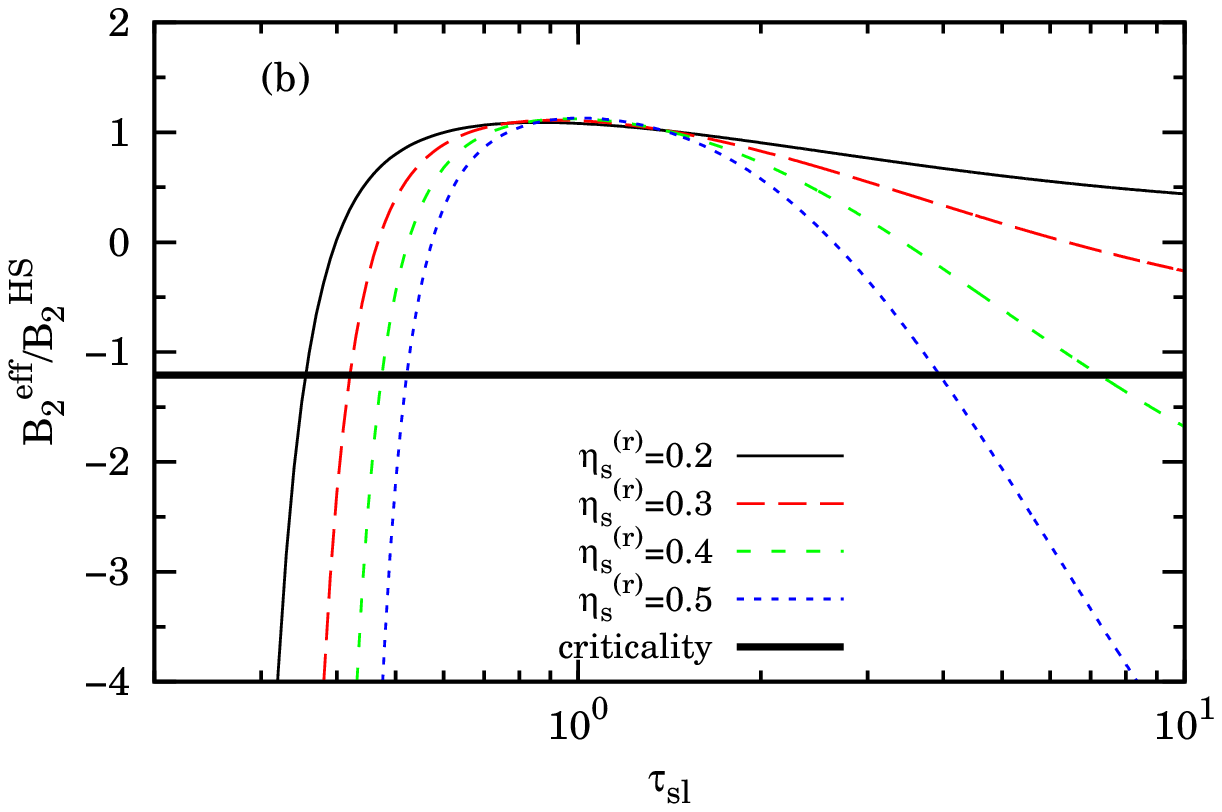}\hfill\includegraphics[width=8cm]{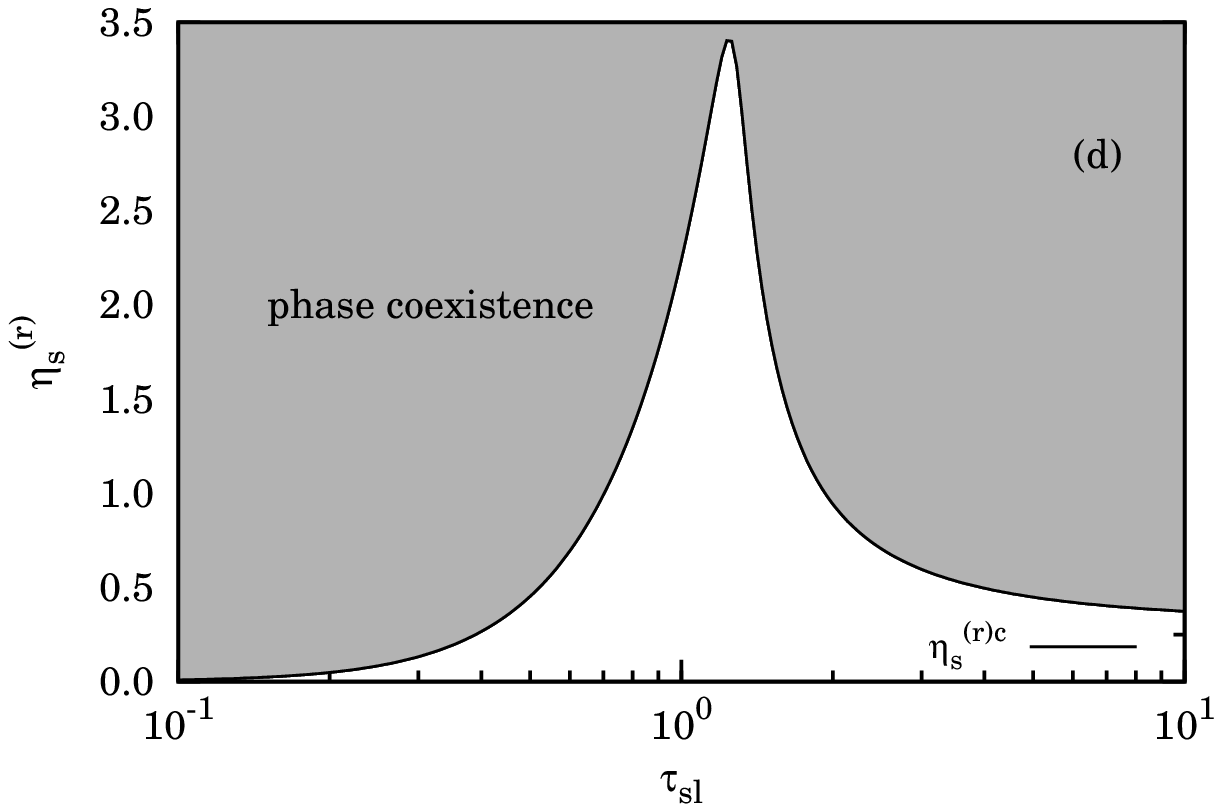}
\caption{Second effective virial coefficient as a
  function of $\tau_{sl}$ for {$q=q_0$} and {several} values of
  $\eta_s^{(r)}$ for {(a)} ASHS and {(b)} SAO models. The thick horizontal line
  corresponds to the {NF} criticality criterion
  $B_2^\text{eff}/B_2^\HS=-1.21$. {Panels (c) and (d) show the dependence of the critical value  $\eta_s^{(r)}=\eta_s^{(r)c}$ (according to the NF criterion) as a function of $\tau_{sl}$ for the ASHS and SAO models, respectively. Note that a logarithmic scale is used on the
  abscissas}. }
\label{fig:b2}
\end{figure*}

It is particularly instructive to observe that {the  NF criterion} confirms the very different critical behavior between the  ASHS model {($\sigma_{ss}/\sigma_s=1$)} and the  SAO model {($\sigma_{ss}/\sigma_s=0$)}.
In {Figs.\ \ref{fig:b2}(a) and \ref{fig:b2}(b)} we compare the second effective virial
coefficients for the two models as  {functions} of $\tau_{sl}$ for
{$q=q_0$} and {several} values of $\eta_s^{(r)}$.
{Here we have identified $\eta_s\to\eta_s^{(r)}$ in the ASHS case, in consistency with the fact that the effective potential is derived in
the infinite solute dilution limit. {The loci of points in the plane $\eta_s^{(r)}$ vs $\tau_{sl}$ where $B_2^\text{eff}/B_2^\HS=-1.21$ are displayed in Figs.\ \ref{fig:b2}(c) and \ref{fig:b2}(d). Inside the shaded regions one has $B_2^\text{eff}/B_2^\HS<-1.21$ and thus phase coexistence is possible, according to the NF criterion.}}

As we already knew from the results of Sec.\ \ref{sec:PYcr}, {Figs.\  \ref{fig:b2}(a) and \ref{fig:b2}(c) show} criticality in the ASHS model only for sufficiently small $\tau_{sl}$ and $\eta_s^{(r)}$.
On the other hand, the scenario present in the SAO model is completely different.
{It is easy to check that  a} critical point in the pure AO
model $(\tau_{sl}\to\infty)$ exists only, {according to the NF criterion}, if $\eta_s^{(r)}\gtrsim 0.318$. {However,}
the presence of stickiness (finite $\tau_{sl}$) {dramatically}
changes the picture. For any $\eta_s^{(r)}$, there exists a critical point if
$\tau_{sl}$ is small enough. Beyond a certain threshold value,
criticality abruptly disappears and then (only if $\eta_s^{(r)}\gtrsim
0.318$) it \emph{re-enters} at a sufficiently large value of
$\tau_{sl}$. Thus, if $\eta_s^{(r)}\gtrsim 0.318$ there exists a window of
values of $\tau_{sl}$ where no phase separation is possible.
{Note that values of $\eta_s^{(r)}>1$, as displayed in Fig.\ \ref{fig:b2}(d), are not unphysical in the SAO model since the reservoir consists in an ideal gas of noninteracting small particles.}

{It must be remarked that} the bridging and pulling effects are more important in the
nonadditive SAO case than in the additive ASHS one, since in the latter the
mutual exclusion of solvent particles interferes with their ability to
attach to the solutes. As illustrated in Fig.\ \ref{fig:b2}, this leads to
paramount differences in the critical behavior of the two {extreme} models.
{For intermediate NASHS models with $0<\sigma_{ss}/\sigma_s<1$ (see Fig.\ \ref{fig:fig3}) a transition from Figs.\ \ref{fig:b2}(b) and \ref{fig:b2}(d) to Figs.\ \ref{fig:b2}(a) and \ref{fig:b2}(c), respectively, can be expected as the excluded volume among the solvent spheres is gradually increased.}

Note also that in the ASHS model {[Fig.\ \ref{fig:b2}(a)]} the results are approximate
(PY) and the solute concentration is zero. On the contrary, in the SAO model
{[Fig.\ \ref{fig:b2}(b)]}, the results are exact and valid for any finite solute
and solvent concentrations. While both models
coincide in the limit of vanishing solvent concentration, in practice
this equivalence requires extremely small values of
$\eta_s^{(r)}$. For instance, at {$\eta_s^{(r)}=10^{-5}$} both values of
$B_2^\eff$ differ by nearly 2\%.
\section{Perturbation theory for the SAO model }
\label{sec:PTII}
From {Sec.\ \ref{sec:NF} we conclude} that the ``hidden''
fluid-fluid phase separation observed by Dijkstra et
al.\cite{Dijkstra99} in their study of the AO model could be
stabilized by adding adhesion, as in our SAO model.
This can be quantified more precisely using a first-order thermodynamic perturbation
theory.\cite{Gast83}

Assuming the HS fluid as reference system, we can write the Helmholtz
free energy per particle {of the effective solute system as}
\beq \label{ptaII}
{\frac{\beta F^\eff}{N_l}=\frac{\beta F_\HS}{N_l}+12\eta_l\eta_s^{(r)}\int_{\sigma_l}^{\sigma_l(1+q)}dr\, r^2\psi(r)g_\HS(r|\eta_l)},
\eeq
where $\beta F_\HS/N_l=(4\eta_l-3\eta_l^2)/(1-\eta_l)^2+\ln(\eta_l)+\text{const}$
is the
Carnahan--Starling\cite{Carnahan69} HS expression, {$\psi(r)$ is given by Eqs.\ \eqref{psidpb}--\eqref{psib}}, and
$g_\HS$ is the HS radial distribution function in the PY approximation\cite{Smith70}, which in the interval $\sigma_l<r<\sigma_l(1+q)<2$ can be written as
\beq
g_\HS(r|\eta_l)=\sum_{n=1}^3\lim_{t\to t_n(\eta_l)}\frac{[t-t_n(\eta_l)]tL(t|\eta_l)}{S(t|\eta_l)}\frac{e^{t(r-1)}}{r},
\label{gHS}
\eeq
where we are measuring lengths in units of $\sigma_l$,
\bq
S(t|\eta_l)&=&(1-\eta_l)^2t^3+6\eta_l(1-\eta_l)t^2+18\eta_l^2t
\nn&&
-12\eta_l(1+2\eta_l),\\
L(t|\eta_l)&=&(1+\eta_l/2)t+1+2\eta_l,
\eq
and $t_n(\eta_l)$ $(n=1,2,3)$ are the zeros of $S(t|\eta_l)$. The first-order Helmholtz
free energy of Eq.\ (\ref{ptaII}) can thus be calculated analytically.

The compressibility factor $Z=\beta p/\rho$ and chemical potential
$\mu$ are then found through
\bq
Z^\eff&=&\eta_l\left.\frac{\partial(\beta
  F^\eff/N_l)}{\partial\eta_l}\right|_{\eta_s^{(r)}},\\
\beta\mu^\eff &=&Z^\eff+\frac{\beta F^\eff}{N_l}.
\eq
The critical point $(\eta_s^{(r)c},\eta_l^c)$ is determined by
numerically solving the following set of equations:
\bq
\left.\frac{\partial (\eta_lZ^\eff)}
{\partial\eta_l}\right|_{\eta_s^{(r)c},\eta_l^c}&=&0,\\
\left.\frac{\partial^2 (\eta_lZ^\eff)}
{\partial\eta_l^2}\right|_{\eta_s^{(r)c},\eta_l^c}&=&0.
\eq

In Fig.\ \ref{fig:cp} we show the critical point
$(\eta_s^{(r)c},\eta_l^c)$ for the fluid-fluid coexistence in {the SAO} model
 {at the threshold value $q=q_0$} as a function of $\tau_{sl}$.
\begin{figure}
\includegraphics[width=8cm]{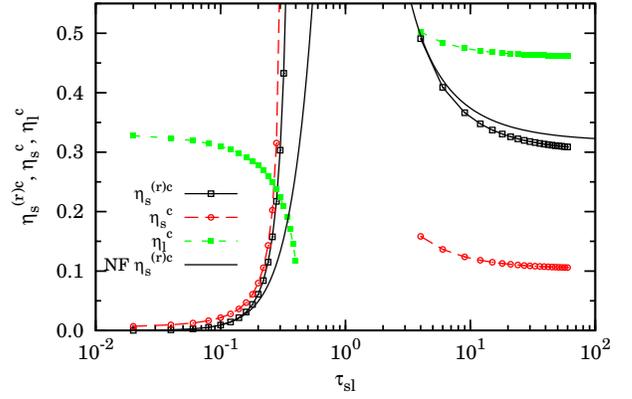}
\caption{Critical point for the fluid-fluid coexistence
  in the {SAO} model  for {$q=q_0$} as a function of $\tau_{sl}$. {The lines with symbols are obtained from perturbation theory, while the solid line corresponds to the NF criterion {[see Fig.\ \protect\ref{fig:b2}(d)]}.} A logarithmic
  scale is used on the abscissa. Equation (\ref{resden2}) is used for the
  conversion between the reservoir  and the solvent
  densities. }
\label{fig:cp}
\end{figure}
The figure confirms the scenario predicted in Sec.\ \ref{sec:NF}
from the NF criterion. {In fact, Fig.\ \ref{fig:cp} shows a relevant mutual consistency between the curves for $\eta_s^{(r)c}$ as obtained from both independent approaches}. There is a range of adhesion for
which there is no {criticality}. For high adhesions {(small $\tau_{sl}$)} we have
{phase} coexistence in the region of low $\eta_s^{(r)}$ region, while for low adhesions {(large $\tau_{sl}$)} the {criticality exists}
in the region of high $\eta_s^{(r)}$.  Of course, we expect a breakdown of the
perturbation theory treatment as soon as {stickiness} becomes too
strong. Also, as soon as $q>q_0$ we are neglecting three-body (and
higher) terms.
\section{Conclusions}
\label{sec:conclusions}

In this paper, we have studied two complementary models of a binary mixture
of (small) solvent and (large) solute particles, where in both cases unlike particles
experience an attractive adhesion interaction of Baxter's type.\cite{Baxter68}
We studied the derivation of an effective solute-solute pair-potential for the two models in
the {regime} of large size asymmetry ($q=\sigma_{s}/\sigma_{l} \ll 1$) and
discussed analogies and differences of the corresponding phase behaviors, as obtained
from the resulting effective one-component fluid.

In the first model, that we dubbed ASHS, both solute-solute and solvent-solvent particles interact as {HSs}
and the reduction to an effective one-component fluid can be carried out only {approximately} via a small {solute} density expansion.
By contrast, this model admits an exact analytical solution within the {PY approximation}.
In the limit of vanishing solute-solvent adhesive attraction, this model reduces to the usual {AHS} binary mixture,
that is known not to display any phase separation within the PY approximation. This {might}, however, be ascribed to the {limitations} of the PY closure, as other more sophisticated theories, as well as numerical simulations,
{support} the existence of phase separation, albeit metastable with respect to freezing, {at} sufficiently large concentrations and size asymmetry {(in this context, nevertheless, see Ref.\ \onlinecite{HTS13})}.
In this case, our analysis {of the ASHS model} confirms previous {findings} of a similar study by Chen et al.\cite{Chen15} in predicting
a closed region in the $(\eta_s,\eta_l)$ plane where phase separation occurs.

While the ASHS model has been around for some time,{\cite{Fantoni05}} the second model {(denoted as SAO) is, to the best of our knowledge, new}.
In this case, solvent particles behave as {an} ideal gas within each other ---but still {they experience a SHS interaction with the solutes}}.
In the limit of no adhesion between solute and solvent, this model reduces to the {well-known} AO one, and we have extended the analysis performed by Dijkstra et al.{\cite{Dijkstra99b}} to the present case.
As in the AO case, even in {the SAO} case the solvent degrees of freedom can be traced out exactly above
a well defined size asymmetry (that is, below a critical value $q_0$ of the size ratio {$q$}), so that the resulting effective one-component {pair}  potential
is \textit{exact}. By contrast, it is not possible in this case to obtain an exact analytical solution {of the binary problem (}not even within the PY approximation{)}, so we
resorted to study a first-order {thermodynamic} perturbation theory of the corresponding exact effective solute-solute pair potential.

In both models, effective potentials can be explained in terms of
``pulling'' and ``bridging'' effects in addition to the usual
``depletion'' mechanism. In the SAO case, the analytical {expressions}
of the effective potential derived in {Eqs.\ (\ref{epaao})--\eqref{psib}} {allow for} an interesting direct physical interpretation. The pulling effect is represented by the term proportional to {$\tau_{sl}^{-1}$ [see Eq.\ \eqref{psip}]}, as the same
(solvent) particle must be in contact with one of the solutes and
outside the exclusion volume of the other solute.
On the
other hand, the bridging effect is represented by the term proportional to {$\tau_{sl}^{-2}$ [see Eq.\ \eqref{psib}]},
as the same (solvent) particle must be in
contact with both solute particles.  These effects are present in both models, but {they}
are more important in the SAO case than in the ASHS case,
since in the latter, the mutual exclusion of solvent particles
interferes with their ability to be attached to the solutes.
{In fact,  the situation sketched in Fig.\ \ref{fig:fig2}(d) is inhibited in the SAO model,  as represented by Fig.\ \ref{fig:b2}(d), which shows always  phase coexistence
at increasing $\eta_s^{(r)}$ for any fixed $\tau_{sl}$.}

{The  derivation of the exact SAO effective potential has allowed us to clearly assess the dramatic influence of solute-solvent attraction on the conventional AO depletion potential. This complements a recent study,\cite{Rovigatti14} where softness in the solute-solvent repulsion was seen to strongly enhance the depletion mechanism.}

Leaving aside the issue of the {metastability with respect to the fluid-solid transition}, the resulting picture confirms the significant impact of
nonadditivity on the {fluid-fluid} phase diagram, as synthesized by Fig.\ \ref{fig:b2}.
Within the {NF  criticality} criterion,
the SAO model is expected to display a {reentrant} phase transition in terms of
$\tau_{sl}$, whereas the ASHS model is not. On the
other hand,  {the} results for the ASHS model are compatible with a
{reentrant} phase transition in terms of $\eta_s$ not observed in the SAO model. A first-order perturbation theory on the SAO model confirms this picture.

Our findings {nicely} confirm {and complement} those by Chen et al.,\cite{Chen15} but extend them
to encompass a \textit{direct} connection with the AO original model, that was missing in the above study,
thus paving the way to a more direct interpretation of the experimental results reported in Refs.\ \onlinecite{Zhao12a} and \onlinecite{Zhao12b}.

While direct numerical simulations of binary mixtures with large size asymmetries are notoriously difficult, it would be interesting to study with numerical experiments whether  adhesion gives rise to the appearance of a metastable fluid-fluid  coexistence
{at large solvent densities for the ASHS model with large $q$ and for the SAO model with very small $q$}.
In addition, they open a number of interesting perspectives for future studies.
Even without resorting to a direct numerical simulation calculations, a number of different {theoretical} approaches can be exploited to make further progresses.

As the attraction between the unlike spheres vanishes ($\tau_{sl} \to \infty$), {the PY solution of} the ASHS model {reduces to the well-known PY solution for a binary AHS mixture,\cite{LR64}  which} does not show phase separation for any size ratio, in spite of the possible depletion
interactions.  {As said above, this seems} to be an artifact of the PY approximation, {as  shown by numerical simulations of the (approximate) effective one-component fluid\cite{Dijkstra99} and by numerical solutions of the Rogers and Young (RY) closure.\cite{Biben91}} Thus, one possibility would be to use the RY closure on a binary mixture with HS interactions between like particles and a short-range {SW} attraction (in the regimes where this can be considered sticky-like\cite{MYS06,Largo08})
between unlike spheres. Work along {these} lines is in progress and will be reported elsewhere.

Another possibility would be to consider a binary {ASHS} mixture with HS interactions between small spheres, weak SHS interactions between the
large spheres, and stronger SHS interactions between small and big spheres.
{This two-component model (which is known to be free from the thermodynamic inconsistency  affecting the one-component model\cite{S91})} could be solved rather easily within the PY approximation, as done for instance by
Zaccarelli et al.\cite{Zaccarelli00}

Finally, it would be nice to extend the {study} reported here for the {ASHS and SAO models} to a more general {NASHS} model where one could tune the solvent-solvent diameter from zero (SAO model)
to the additive value (ASHS model), thus encompassing both models into an unified framework.
{MC simulations for a binary ASHS mixture have been performed by Jamnik,\cite{Jamnik08}
but not for the determination of the phase diagram, which has been
studied for the one-component SHS fluid by Miller and Frenkel.\cite{Miller03,Miller04} To the best of our knowledge, no
numerical experiment has ever been tried on the NASHS
binary mixture.}

\begin{acknowledgments}
A.G. gratefully acknowledges support from PRIN-MIUR 2010-2011 project (contract 2010LKE4CC). The research of A.S. has been partially supported by the Spanish Government through Grant No.\ FIS2013-42840-P and by the Regional Government of Extremadura (Spain) through Grant No.\ GR15104 (partially financed by ERDF funds).

\end{acknowledgments}

\appendix
\section{{A} simple geometrical argument related to Fig.\ \protect\ref{fig:fig2}}
\label{app:simple}
We first estimate how many small spheres of diameter $\sigma_s$ are necessary
to cover the surface of a large sphere of diameter $\sigma_l$. Assuming $q=\sigma_s/\sigma_l \ll 1$,
the small spheres will be distributed on the large sphere surface approximately with
a hexagonal packing {corresponding to an \emph{area} fraction} $\eta_{\text{hex}}=\pi/2 \sqrt{3} \approx {0.907}$.
{Thus, $\eta_{\text{hex}}=\phi a/A$, where $\phi$ is the number of the adsorbed small spheres, $a=(\pi/4){\sigma_s^2}$ is the area of the projected disk associated with each solvent sphere, and  $A=\pi\sigma_l^2$ is the surface area of the solute particle. Therefore,}
\beq
\label{appa:eq1}
{\phi=\eta_{\text{hex}}\frac{A}{a}= \frac{2\pi}{\sqrt{3}} q^{-2}}.
\eeq
The critical volume fraction $\eta_s^{*}$ at which all large colloidal spheres, distributed with
a volume fraction $\eta_l=(\pi/6) \rho_l \sigma_l^3$, {can} be covered {is}
\begin{eqnarray}
\label{appa:eq2}
\eta_s^{*}&=& \eta_l {q^{3}} \phi
\end{eqnarray}
and this leads to the expression reported in Sec.\ \ref{sec1}.
\section{Analytical PY expressions for the ASHS model {in the limit $x_l\to 0$} }
\label{app:RFA}
{The Rational-Function Approximation (RFA) methodology\cite{Santos98,Yuste08,HYS08}
is known to  give access to
analytical formulae of the PY solution for the ASHS model.\cite{Baxter68,Baxter70,Fantoni05}
In this Appendix we assume the infinite dilution limit for the solutes ($x_l\to 0$).}

{According to Eq.\ (36) of Ref.\ \onlinecite{Yuste08}, the Laplace transform $G_{ll}(s)=\int_0^\infty dr\, e^{-sr}r g_{ll}(r)$ of $rg_{ll}(r)$ is, in the limit $x_l\to 0$,}
\beq
G_{ll}(s)=\frac{e^{-s}}{s^2}\left[L_{ll}(s)+L_{ls}(s)\frac{A_{sl}(s)}{1-A_{ss}(s)}\right],
\label{6}
\eeq
{where $\sigma_l=1$ has been chosen as length unit and\cite{Santos98}}
\beq
L_{ll}(s)=L_{ll}^{(0)}+L_{ll}^{(1)}s,
\label{7}
\eeq
\beq
L_{ls}(s)=L_{ls}^{(0)}+L_{ls}^{(1)}s+L_{ls}^{(2)}s^2,
\label{8}
\eeq
\beq
A_{sl}(s)={12\eta_s}\left[\phi_2(qs)L_{sl}^{(0)}+\frac{\phi_1(qs)}{q}L_{sl}^{(1)}+\frac{\phi_0(qs)}{q^2}L_{sl}^{(2)}\right],
\label{9}
\eeq
\beq
A_{ss}(s)={12\eta_s}\left[\phi_2(qs)L_{ss}^{(0)}+\frac{\phi_1(qs)}{q}L_{ss}^{(1)}\right].
\label{10}
\eeq
{Here, $\phi_n(x)\equiv -x^{-(n+1)}\left[e^{-x}-\sum_{j=0}^n (-x)^j/j!\right]$.
The coefficients $L_{\alpha\gamma}^{(k)}$ are given by\cite{Santos98}}
\beq
L_{ll}^{(0)}=L_{sl}^{(0)}=\frac{1-({12\eta_s}/{ q^2})L_{sl}^{(2)}}{1-\eta_s}+\frac{3\eta_s}{q(1-\eta_s)^2},
\label{12}
\eeq
\beq
L_{ls}^{(0)}=L_{ss}^{(0)}=\frac{1}{1-\eta_s}+\frac{3\eta_s}{(1-\eta_s)^2},
\label{13}
\eeq
\beq
L_{ll}^{(1)}=\frac{1-({6\eta_s}/{ q^2})L_{sl}^{(2)}}{1-\eta_s}+\frac{3\eta_s}{2q(1-\eta_s)^2},
\label{14}
\eeq
\beq
L_{sl}^{(1)}=\frac{1+q-({12\eta_s}/{ q})L_{sl}^{(2)}}{2(1-\eta_s)}+\frac{3\eta_s}{2(1-\eta_s)^2},
\label{15}
\eeq
\beq
L_{ls}^{(1)}=\frac{1+q}{2(1-\eta_s)}+\frac{3\eta_s}{2(1-\eta_s)^2},
\label{16}
\eeq
\beq
L_{ss}^{(1)}=q \frac{1+\eta/2}{(1-\eta)^2},
\label{17}
\eeq
\beq
L_{ls}^{(2)}=L_{sl}^{(2)}=\frac{1}{12}\frac{1+q+\frac{3\eta_s}{1-\eta_s}}{4\tau_{sl}\frac{1-\eta_s}{1+q}+{\eta_s}/{q}}.
\label{18}
\eeq

This closes the determination of $G_{ll}(s)$ for given values of $\eta_s$, $q$, and $\tau_{sl}$. Then, by numerical inverse transform one can easily obtain $g_{ll}(r)$.  On
the other hand, pure analytical expressions are also possible for the
different layers $1<r<1+q$, $1+q<r<1+2q$, $1+2q<r<1+3q$, \ldots. The
trick consists in formally attaching a bookkeeping factor $\varepsilon$
to any exponential in $G_{ll}(s)$. Then, by expanding in powers of
$\varepsilon$ we can write
\bq
G_{ll}(s)=\sum_{n=0}^\infty e^{-(1+nq)s}\Gamma_n(s),
\label{19}
\eq
where we have made $\varepsilon=1$. From Eq.\ \eqref{19} we get
\bq
g_{ll}(r)=\frac{1}{r}\sum_{n=0}^\infty \Theta(r-1-nq)\gamma_n(r-1-nq),
\label{20}
\eq
where $\gamma_n(r)$ is the inverse Laplace transform of
$\Gamma_n(s)$. The functions $\gamma_n(r)$ can then be expressed
in terms of the three roots of a cubic equation, {analogously to the case of Eq.\ \eqref{gHS}}. Therefore, if we
are only interested in the interval $1\leq r\leq 1+kq$, we just need to keep
the first $k$ terms in the sum of Eq.\ (\ref{20}).

{From a practical point of view, it is sufficient to determine $g_{ll}(r)$ in the interval $1\leq r \leq 1+3q$, in which case only $\gamma_0(r)$, $\gamma_1(r)$, and $\gamma_2(r)$ are needed. Their analytical expressions are easily obtained with a computational software program but are too lengthy to be reproduced here. In general, $\gamma_1(0)\neq 0$, what implies a jump discontinuity of $g_{ll}(r)$ at $r=1+q$,}
\beqa
{\delta g_{ll}}&\equiv&{g_{ll}((1+q)^-)-g_{ll}((1+q)^+)}={-\frac{\gamma_1(0)}{1+q}}\nn
&=&{\frac{(1+q)\eta_s\left[1+q+(2-q)\eta_s\right]^2}{12q(1-\eta_s)^2\left[(1+q)\eta_s+4q\tau_{sl}(1-\eta_s)\right]^2}}.\nn
&&
\label{B15}
\eeqa
{Note that $r=1+q$ is the threshold distance beyond which no bridges are possible (see Fig.\ \ref{fig:fig2}). This is clearly reflected by  a strong decrease of $g_{ll}(r)$ when going from $r=(1+q)^-$ (bridges are possible) to $r=(1+q)^+$  (no bridging effect). This physical phenomenon can give rise, as an artifact of the PY approximation, to a negative value of $g_{ll}(r)$ at $r=(1+q)^+$ if $\eta_s$ is sufficiently large or $\tau_{sl}$ is sufficiently small. This is illustrated in Fig.\ \ref{fig:locus} for {$q=0.12$ and $q=q_0$.}}

\begin{figure}
\includegraphics[width=8cm]{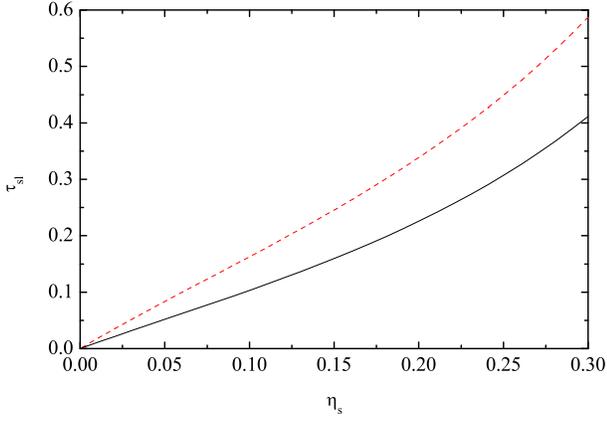}
\caption{{Loci in the plane $\tau_{sl}$ vs $\eta_s$ where the PY approximation predicts $g_{ll}(r)=0$ at $r=(1+q)^+$ for  {$q=0.12$} (upper curve) and {$q=q_0$} (lower curve). The radial distribution function $g_{ll}(r)$ is not positive definite  below each curve.}
\label{fig:locus}}
\end{figure}

{Once $g_{ll}(r)$ is known, Eq.\ \eqref{vll} gives the pair potential $v_{ll}(r)$ in the PY approximation, as depicted in Fig.\ \ref{fig:potential}.}
The  effective second virial coefficient can also be determined
analytically as follows:
\begin{widetext}
\beqa
\label{B2ASHS}
{B_2^\eff}&=&-2\pi\int_0^\infty dr\, r^2\left[g_{ll}(r)-1\right]=2\pi\lim_{s\to 0}\frac{\partial}{\partial s}\left[G_{ll}(s)-s^{-2}\right]\nn
&=&
{\frac{\pi}{12(1+2\eta_s)^2}\Bigg\{
8 + (20 - 15 q - 6 q^2 - q^3) \eta_s +
 2 (4 - 6 q + 3 q^2 + q^3) \eta_s^2 - q^3 \eta_s^3}\nn
 &&
   { +\frac{2 (1 + q)\eta_s (1 + q + 2 \eta_s - q \eta_s) \left[6 (1 + 2 \eta_s) +
    q^2 (1 - \eta_s)^2 +
    2 q (1 - \eta_s) (2 + \eta_s)\right]}{(1 - \eta_s) \left[(1 + q) \eta_s +
    4 q \tau_{sl} (1 - \eta_s)\right]}}\nn
 &&{- \frac{(1 + q)^2 \eta_s (1 + q + 2 \eta_s -
     q \eta_s)^2(2 + q  + 4 \eta_s-q\eta_s)  }{(1 - \eta_s) \left[(1 + q) \eta_s +
    4 q \tau_{sl} (1 - \eta_s)\right]^2}\Bigg\}}   .
 \label{B16}
\eeqa
\end{widetext}
\section{{Derivation} of the relationship between $\eta_s$, {$\eta_{l}$,} and $\eta_s^{(r)}$
in the SAO model}
\label{app:alternative}
{In the semi-grand-canonical ensemble $(z_s,N_l,V,T)$ the average number of small particles can be obtained from the associated thermodynamic potential $\mathcal{F}$ as }
\beq
\langle N_s\rangle_{z_s,N_l}=-z_s\frac{\partial\beta\mathcal{F}}{\partial z_s}.
\label{C1}
\eeq
{Now, from Eq.\ \eqref{general:eq10} and the equality $U_{ll}^\eff=U_{ll}+\Omega$, we can write}
\beq
e^{-\beta\mathcal{F}}=\left\langle e^{-\beta\Omega}\right\rangle_{N_l}\frac{\int d\rr^{N_l}\, e^{-\beta U_{ll}}}{N_l!\Lambda_l^{3N_l}},
\eeq
{where}
\beq
\left\langle\cdots\right\rangle_{N_l}=\frac{\int d\rr^{N_l}\, \cdots e^{-\beta U_{ll}}}{\int d\rr^{N_l}\, e^{-\beta U_{ll}}}
\eeq
denotes a canonical average over the {bare} solutes.
{Then, taking into account that $\Omega\propto z_s$, Eq.\ \eqref{C1} reduces to}
\beq
\langle N_s\rangle_{z_s,N_l}=-\frac{\left\langle e^{-\beta\Omega}\beta\Omega\right\rangle_{N_l}}{\left\langle e^{-\beta\Omega}\right\rangle_{N_l}}.
\eeq
{Next, if $q<q_0$, $\Omega=\Omega_0+\Omega_1+\Omega_2$, so that}
\beq
\langle N_s\rangle_{z_s,N_l}=-\beta\Omega_0-\beta\Omega_1-\frac{\left\langle e^{-\beta\Omega_2}\beta\Omega_2\right\rangle_{N_l}}{\left\langle e^{-\beta\Omega_2}\right\rangle_{N_l}}.
\label{C5}
\eeq
{Note that the last term on the right-hand side can be rewritten as}
\beq
\frac{\left\langle e^{-\beta\Omega_2}\beta\Omega_2\right\rangle_{N_l}}{\left\langle e^{-\beta\Omega_2}\right\rangle_{N_l}}=\frac{\rho_l^2}{2}V\int d\rr \,\beta v_{ll}(r)g_{\eff}(r|\eta_l,\eta_s^{(r)}),
\label{C6}
\eeq
{where}
\beq
g_{\eff}(r_{12}^{(l)}|\eta_l,\eta_s^{(r)})=\frac{V^2\int d\rr_3^{(l)}\cdots \int d\rr_{N_l}^{(l)}e^{-\beta U_{ll}^\eff}}
{\int d\rr^{N_l}e^{-\beta U_{ll}^\eff}}
\eeq
{and we have taken into account that  $N_l(N_l-1)\simeq N_l^2$ in the thermodynamic limit.}
{Finally, applying Eqs.\ \eqref{4.7}, \eqref{4.11}, and \eqref{C6} in Eq.\ \eqref{C5}, it is easy to obtain Eq.\ \eqref{resden}.}

{The first two terms on the right-hand side of Eq.\ \eqref{resden}  can also be obtained from the canonical ensemble $(N_s,N_l,V,T)$. Up to the level of the second virial coefficient, the free energy $F$ is}
\beqa
\frac{\beta F}{V}&=&\rho_s\ln\left(\rho_s\Lambda_s^3\right)+\rho_l\ln\left(\rho_l\Lambda_l^3\right)-\rho_s-\rho_l+\frac{2\pi}{3}\rho_l^2\nn
&&+2B_{sl}\rho_s\rho_l+\mathcal{O}(\rho^3),
\eeqa
{where $B_{sl}=\frac{\pi}{12}\sigma_l^3(1+q)^3\left(1-\frac{1}{4\tau_{sl}}\right)$. The solvent chemical potential is $\mu_s={\left[\partial (F/V)/\partial \rho_s\right]}_{\rho_l}$, so that}
\beq
z_s\equiv\frac{e^{\beta\mu_s}}{\Lambda_s^3}=\rho_s\left[1+2B_{sl}\rho_l+
{\cal O}(\rho^2)\right], \label{resdenapp}
\eeq
which is consistent with Eq.\ (\ref{resden}).

\section{SWAO model}
\label{app:SWAO}
{In the SWAO model Eq.\ \eqref{general:eq3b} is replaced by}
\bq
f_{sl}(r)=\left\{\begin{array}{ll}
-1, & r<\sigma_{sl},\\
e^{\beta\epsilon_{sl}}-1, & \sigma_{sl}<r<\sigma_{sl}(1+\Delta_{sl}),\\
0, & r>\sigma_{sl}(1+\Delta_{sl}),
\end{array}\right.
\label{D1}
\eq
{where $\epsilon_{sl}$ and $\sigma_{sl}\Delta_{sl}$ are the depth and width, respectively, of the attractive well. One can define an effective stickiness parameter\cite{MYS06} $\tau_{sl}^{-1}=12\left(e^{\beta\epsilon_{sl}}-1\right)\Delta_{sl}$, so that the SWAO model reduces to the SAO one in the double limit $\epsilon_{sl}\to\infty$, $\Delta_{sl}\to 0$ at fixed $\tau_{sl}$.}

{All the steps in Sec.\ \ref{sec:AAO} up to Eq.\ \eqref{vllff} are still valid for the SWAO model. However, the condition for having $\Omega_n=0$ if $n\geq 3$ is not $\sigma_{sl}<\sigma_l(1+q_0)/2$ (or $q<q_0$) but $\sigma_{sl}(1+\Delta_{sl})<\sigma_l(1+q_0)/2$, what is equivalent to $q(1+\Delta_{sl})+\Delta_{sl}<q_0$.}

{To simplify the expressions, in this appendix we take again $\sigma_l=1$ as the length unit. Inserting Eq.\ \eqref{D1} into Eq.\ \eqref{vllff}, one obtains}
\beq
\beta v_{ll}(r)=\eta_s^{(r)}\begin{cases}
  \infty,&r<1,\\
    {\psi(r)},&1<r<(1+q)\left(1+\Delta_{sl}\right),\\
  0,&r>(1+q)\left(1+\Delta_{sl}\right),
\end{cases}
\label{vllSWAO}
\eeq
{where the function $\psi(r)$ can again be decomposed into three terms (depletion$+$pulling$+$bridging), as given by Eq.\ \eqref{psidpb}, except that now}
\beqa
{\psi_{\text{p}}(r)}&=&{\frac{(\tau_{sl}\Delta_{sl})^{-1}}{8\pi q^3}\left[\mathcal{C}(r_-,1+q)+\mathcal{C}(r_+,(1+q)(1+\Delta_{sl}){)}\right.}\nn
&&{\left.-2\mathcal{C}(r,1+q)\right]},
\eeqa
\beqa
{\psi_{\text{b}}(r)}&=&{-\frac{(\tau_{sl}\Delta_{sl})^{-2}}{96\pi q^3}\left[\mathcal{C}(r,1+q)+\mathcal{C}(r,(1+q)(1+\Delta_{sl}))\right.}\nn
&&{\left.-\mathcal{C}(r_-,1+q)-\mathcal{C}(r_+,(1+q)(1+\Delta_{sl}))\right]},
\eeqa
{where}
\beq
{\mathcal{C}(r,a)=\frac{\pi}{3}(a-r)^2(2a+r)\Theta(a-r)}
\eeq
{is the volume of a spherical cap of height $a-r$ in a sphere of radius $a$ and}
\beq
{r_{\pm}\equiv r\pm\frac{(1+q)^2}{2r}\Delta_{sl}\left(1+\frac{\Delta_{sl}}{2}\right).}
\eeq
{The depletion term is still given by Eq.\ \eqref{psid}, i.e.,
$\psi_{\text{d}}(r)=-(3/2\pi q^3)\mathcal{C}(r,1+q)$.}

{The ranges of the contributions $\psi_{\text{d}}(r)$, $\psi_{\text{p}}(r)$, and $\psi_{\text{b}}(r)$ are $1+q$, $(1+q)(1+\Delta_{sl}/2)$, and $(1+q)(1+\Delta_{sl})$, respectively.} It can be easily verified that in the
sticky limit $\Delta_{sl}\to 0$ the potential of Eq.\ (\ref{vllSWAO})
reduces to the one of Eq.\ (\ref{epaao}). One can also verify that the
jump discontinuity at $r=2\sigma_{sl}{=1+q}$ of the {SAO} model disappears in
the SWAO one, which is everywhere continuous.


\begin{thebibliography}{53}
\bibitem[{not()}]{note_1}
\bibinfo{note}{Usually, when talking about the AO model one refers to a
  colloid-polymer mixture where the depletants are linear homopolymers of
  radius of gyration $\sigma_s/2$ in a good solvent. After tracing out the
  monomer degrees of freedom and replacing each chain with a particle at its
  center of mass, the homopolymers can be considered, to a first level of
  approximation, as noninteracting among themselves but unable to penetrate a
  sphere of diameter $\sigma_l + \sigma_s$ around each colloidal particle. In
  this work, however, we will rather talk about a solvent-solute mixture.}
\end{thebibliography}
\end{document}